\documentclass[twocolumn,twocolappendix]{aastex7}

\definecolorset{HTML}{CL}{}{Rosewater,DC8A78;Flamingo,DD7878;Pink,EA76CB;Mauve,8839EF;Red,D20F39;Maroon,E64553;Peach,FE640B;Yellow,DF8E1D;Green,40A02B;Teal,179299;Sky,04A5E5;Sapphire,209FB5;Blue,1E66F5;Lavender,7287FD;Text,4C4F69;Subtext1,5C5F77;Subtext0,6C6F85;Overlay2,7C7F93;Overlay1,8C8FA1;Overlay0,9CA0B0;Surface2,ACB0BE;Surface1,BCC0CC;Surface0,CCD0DA;Base,EFF1F5;Mantle,E6E9EF;Crust,DCE0E8}
\hypersetup{linkcolor=CLRed,citecolor=CLBlue,filecolor=CLSky,urlcolor=CLMauve}

\usepackage[T1]{fontenc}
\usepackage{amsmath}  
\usepackage{bm}  
\usepackage{newunicodechar}

\usepackage{xfrac}\usepackage{anyfontsize}  
\usepackage{booktabs}   
\usepackage[capitalise,noabbrev]{cleveref}  
\usepackage{enumitem}
\usepackage{etoolbox}
\usepackage{fancyvrb}\VerbatimFootnotes{}  

\makeatletter\pretocmd\@sect{\def\@currentcounter{#1}}{}{\fail}\makeatother

\graphicspath{{./}{}}

\usepackage{mathtools}

\usepackage{siunitx}  
\DeclareSIUnit{\hubble}{\mathnormal{h}}  
\DeclareSIUnit{\parsec}{pc}
\DeclareSIUnit{\angstrom}{\Angstrom}

\usepackage[acronym]{glossaries-extra}
\setabbreviationstyle[acronym]{long-short}
\glsdisablehyper{}  
\newacronym{2lpt}{2LPT}{second-order Lagrangian perturbation theory}
\newacronym{2pcf}{2PCF}{two-point correlation function}
\newacronym[
longplural={baryon acoustic oscillations}
shortplural={BAO}
]{bao}{BAO}{baryon acoustic oscillations}
\newacronym{cdm}{CDM}{cold dark matter}
\newacronym{cic}{CIC}{cloud-in-cell}
\newacronym{de}{DE}{dark energy}
\newacronym{dm}{DM}{dark matter}
\newacronym{dtfe}{DTFE}{Delaunay tesselation field estimator}
\newacronym{fof}{FOF}{friends-of-friends}
\newacronym{ic}{IC}{initial condition}
\newacronym{lpt}{LPT}{Lagrangian perturbation theory}
\newacronym{lss}{LSS}{large-scale structure}
\newacronym{msc}{MSC}{Morse-Smale complex}
\newacronym{mse}{MSE}{Morse-Smale complex extraction}
\newacronym{ph}{PH}{persistent homology}
\newacronym{pm}{PM}{particle-mesh}
\newacronym{tda}{TDA}{topological data analysis}
\newacronym{tpm}{TreePM}{tree-particle-mesh}
\newacronym{za}{ZA}{Zel'dovich approximation}
\newacronym{nh}{NH}{normal hierarchy}
\newacronym{ih}{IH}{inverted hierarchy} 
\newacronym{cmb}{CMB}{cosmic microwave background}
\newacronym{dr1}{DR1}{Data Release 1}
\newacronym{dr2}{DR2}{Data Release 2}
\newacronym{dr3}{DR3}{Data Release 3}

\newcommand{\DisPerSE}{\textsc{DisPerSE}}
\newcommand{\Gadget}[1][III]{\ifstrempty{#1}{\textsc{Gadget}}{\textsc{Gadget-#1}}}
\newcommand{\Pylians}{\textsc{Pylians}}
\newcommand{\Quijote}{\textsc{Quijote}} 
\newcommand{\Camb}{\textsc{CAMB}}

\received{2026 April 08}
\submitjournal{\apj}

\begin{document}


\title{Signatures of Massive Neutrinos in the Cosmic Web via Persistent Homology}

\shorttitle{Persistent Homology and Massive Neutrinos}
\shortauthors{Yu et al. (2026)}

\author[sname=Yu,gname=Hogyun]{Hogyun Yu}
\affiliation{Department of Physics and Astronomy\\
  Sejong University, Seoul, 143--747, Republic of Korea}  
\email{...@...}  

\author[orcid=0000-0002-2697-2831,sname=Michaux,gname=Michaël]{Micha{\"e}l Michaux}  
\affiliation{Department of Physics and Astronomy\\
  Sejong University, Seoul, 143--747, Republic of Korea}
\email{...@...}  

\author[sname=Kim,gname=Donghyun]{Donghyun Kim}  
\affiliation{Department of Physics and Astronomy\\
  Sejong University, Seoul, 143--747, Republic of Korea}
\email{...@...}  
  
\author[sname=Song,gname=Changhee]{Changhee Song}  
\affiliation{Department of Physics and Astronomy\\
  Sejong University, Seoul, 143--747, Republic of Korea}
\email{...@...}  

\author[sname=Yoon,gname=Ingyu]{Ingyu Yun}  
\affiliation{Department of Physics and Astronomy\\
  Sejong University, Seoul, 143--747, Republic of Korea}
\email{...@...}  

\author[sname=Lee,gname=Donghyeon]{Donghyeon Lee}  
\affiliation{Department of Physics and Astronomy\\
  Sejong University, Seoul, 143--747, Republic of Korea}
\email{...@...}  

\author[sname=Lee,gname=Yoonyoung]{Yoonyoung Lee}  
\affiliation{Department of Physics and Astronomy\\
  Sejong University, Seoul, 143--747, Republic of Korea}
\email{...@...}  

\author[sname=Rossi,gname=Graziano]{Graziano Rossi}  
\affiliation{Department of Physics and Astronomy\\
  Sejong University, Seoul, 143--747, Republic of Korea}
\email[show]{graziano@sejong.ac.kr}

\correspondingauthor{Graziano Rossi}


\begin{abstract}

We present the second paper in our program characterizing the impact of massive neutrinos on 
the \textit{multiscale} cosmic web using global topology and persistent homology. 
Building on the methodology established in Paper I, based on discrete Morse theory, 
we analyze a subset of the \Quijote{} simulations to compute 
persistent diagrams, Betti curves, and additional topological statistics 
for both dark matter  
and halo density fields, across redshifts  $z=0,1,2$.   
A central result of our study is the first clear demonstration that apex points in persistent 
diagrams are especially sensitive to neutrino mass, 
with enhanced sensitivity for specific pairs of saddle points at high redshift. 
In addition, Betti curves from dark matter  
density fields broaden and flatten with increasing neutrino masses, exhibiting two characteristic
density thresholds where Betti numbers remain invariant. 
These mass-dependent signatures are detectable at the few-percent level, even for $M_\nu \sim 0.1$ eV, providing a 
robust, physically grounded probe of massive neutrinos in the cosmic web. 
While traditional two-point statistics encode only pairwise correlations and cannot fully break parameter degeneracies, 
persistent homology captures higher-order, \textit{multiscale} information that can lift these degeneracies.
Moreover, its high sensitivity to the sum of neutrino masses makes it a promising complement to conventional analyses. 
Our results thus establish a solid foundation for
forward-modeling  or emulator-based approaches using persistent homology and environment-based 
statistics to constrain neutrino mass---potentially enabling direct detection---and additional cosmological parameters,
with immediate relevance for ongoing and upcoming galaxy surveys, including DESI, Euclid, and Rubin-LSST.

\end{abstract}

 

\section{Introduction}      \label{sec:introduction}

The latest cosmological results from the Dark Energy Spectroscopic Instrument \citep[DESI;][]{DESI:2016} 
are challenging the $\Lambda$CDM framework---the concordance, spatially flat model dominated by \gls{cdm} and a cosmological 
constant ($\Lambda$) \gls{de} component---revealing intriguing deviations from the standard scenario, including hints of 
dynamical \gls{de} and unexpectedly stringent constraints on the summed neutrino mass $M_\nu$ \citep{DESI-DR1-Cosmo,DESI-DR2-Cosmo}. 
In particular, the most up-to-date analyses combining DESI \gls{dr2} with \gls{cmb} 
measurements report $M_\nu < 0.0642~\mathrm{eV}$ within the $\Lambda$\gls{cdm} 
framework \citep{DESI-DR2-Cosmo}. 
Moreover, in settings with an unconstrained prior on $M_\nu$, 
results approach---or formally violate---the minimal mass allowed for the neutrino \gls{nh}, and in certain cases yield 
unphysical $M_\nu$ values. 
These findings  have reignited debate in the literature, including speculative proposals such as the contentious idea of 
``negative neutrino mass'' \citep[e.g.,][]{Naredo-Tuero2024,GreenMeyers2025}. 
Additional data from the final DESI five-year observations \gls{dr3}, Euclid \citep{Laureijs2011}, and next-generation surveys 
such as the Rubin Observatory Legacy Survey of Space and Time \citep[LSST;][]{LSST:2019} and the 
Nancy Grace Roman Space Telescope \citep[Roman;][]{Spergel2015} may help clarify these issues. 
Meanwhile, cosmology remains at a puzzling juncture, with current results raising fundamental questions about the standard model.

Rather than invoking nonstandard physics or exotic scenarios, a promising alternative approach stems from recognizing 
that traditional procedures for cosmological parameter estimation and neutrino mass constraints 
typically rely on two-point summary statistics, 
such as the total matter power spectrum or the two-point correlation function. 
Moreover, analyses based on these conventional methods often focus on individual tracers---galaxies, clusters, or voids---at selected scales. 
Yet, two-point statistics compress the density field, capturing only pairwise correlations 
while discarding higher-order and non-Gaussian topological information. 
At the same time, the cosmic web is intrinsically \textit{multiscale},
exhibiting complex structures across a wide range of 
densities and scales \citep[e.g.,][]{Zel'dovich:1970,Shandarin:1989,Bond:1996,AragonCalvo:2010,DisPerSE:2011a,Cautun2014,
Feldbrugge2019,Pranav2019,Wilding2021,AragonCalvo:2024}. 
This motivates the development of statistical approaches that go beyond conventional two-point measures, 
explicitly exploiting the full \textit{multiscale} topology and connectivity of the cosmic web 
rather than isolated components or limited scales,
capturing its full complexity. 
Leveraging this \textit{multiscale} structure via higher-order 
methods---particularly to probe neutrino mass---is the central objective of the present work,
building upon our foundational studies \citep{Rossi:2022,Rossi:2026}.

To this end, a powerful approach to characterizing the \textit{multiscale} cosmic web is provided by 
Discrete Morse theory \citep[see pioneering contributions by][]{Forman1998,Forman2002} 
and persistent homology, first introduced by \cite{Edelsbrunner2000}, \cite{Edelsbrunner2002}, and \cite{Robins1998,Robins2000}, which constitute 
core methods in computational topology \citep{Gyulassy2008,Zomorodian2009} and \gls{tda}. These techniques overcome the limitations of 
classical Morse theory \citep[e.g.,][]{Milnor1963} when applied to noisy, irregular, or discrete datasets, enabling robust identification of 
genuine topological structures. Spurious features are removed via a \textit{filtration} procedure, or \textit{topological simplification}, 
ensuring that persistence pairs represent meaningful topological components.

Such methods have wide-ranging applications across scientific disciplines and are increasingly 
adopted in cosmology to probe the \textit{multiscale} cosmic web. 
In this context, they establish a direct correspondence between critical points of the 
density field and cosmic web components---voids, walls, filaments, and clusters---allowing for a hierarchical, 
\textit{multiscale} characterization that is robust to sampling effects and sensitive to underlying physical differences \citep[see review in][]{Rossi:2026}. 
Key recent contributions include
\cite{Feldbrugge2019},  \cite{Pranav2019}, \cite{Pranav2021}, 
\cite{ Wilding2021}, \cite{Biagetti2022}, \cite{Heydenreich2022}, \cite{Calles2025}.

Importantly, higher-order, nonlinear \textit{multiscale} features of the cosmic web are sensitive to fundamental physical processes, 
including the effects of massive neutrinos.  
However, while neutrino impacts on cosmological perturbations are well understood at linear order \citep[e.g.,][]{LesgourguesPastor2006,Rossi2017}, 
linear theory cannot capture key aspects of structure formation in the nonlinear regime, where baryonic physics, mode coupling, and small-scale clustering become relevant. 
In this regime, neutrino effects are more subtle and comparatively less investigated.

Existing studies have explored neutrino effects across diverse observables and scales  
\citep[e.g.,][]{Ajani2020,Bolliet2020,Hahn2020,Kuruvilla2020,Rossi2017,Rossi2020,Zhang2020,Bose2021,Massara2021,Whitford2022,Verza2023,Yankelevich2023,Cueli2024,Thiele2024,Labate2025,Luchina2025,Maggiore2025}, 
yet most rely on a single probe or limited range of scales, thereby compressing the  rich \textit{multiscale} information encoded in the cosmic web.

In reality, neutrino imprints are inherently \textit{multiscale}: by modifying the growth of density fluctuations across scales, 
they leave subtle but detectable signatures on the topology and connectivity of the matter distribution. 
This makes the cosmic web 
an ideal target for persistent homology and \gls{tda}. Higher-order statistics can therefore 
capture the full complexity of cosmic structures and their 
dependence on neutrino mass.

To this end, \citet{Rossi:2022} were the first to demonstrate the relevance of persistent homology in massive-neutrino cosmologies, 
with \citet{Moon:2023} extending the analysis to critical points, their clustering, and large-scale features such as \gls{bao} amplitudes and inflection scales. 
\citet{Rossi:2026} further applied  
discrete Morse theory, global topology, and persistent homology to characterize the \textit{multiscale} 
cosmic web (with a special focus on filaments), highlighting the potential of topological methods and \gls{tda} 
as precision probes of neutrino mass and cosmological parameter constraints. 
These works established the physical relevance of persistence-based techniques and motivate further development, 
pursued in this study and within our long-term research program.
 
Notably, despite these advances, topological methods remain comparatively 
underexplored in massive-neutrino cosmologies \citep[e.g.,][]{JalaliKanafi:2024,Yip2024,Prat2026}, 
with only recently more sophisticated higher-order techniques beginning to emerge as observational data improve.
Our long-term program \citep[e.g.,][]{Rossi:2026} is designed to fill this gap, leveraging the full higher-order \textit{multiscale} 
information of the cosmic web for precision studies of massive neutrinos
and to advance neutrino mass constraints.

In the present study, building on our seminal works 
\citep{Rossi:2022,Moon:2023,Rossi:2026}, we compute persistence diagrams, 
Betti curves, and related topological statistics from a selected sample of \Quijote{} 
$N$-body simulations \citep{Quijote:2020}.
We analyze both the \gls{dm} density field on a regular grid, following an approach similar to \citet{Moon:2023}, 
and the corresponding halo catalogs, following the methodology of \citet{Rossi:2026} via Delaunay triangulation \citep{DisPerSE:2011a,DisPerSE:2011b}:
our methodology is summarized in \Cref{sec:methodology}. 
We consider three cosmologies with identical initial conditions and random seeds---a fiducial Planck-like massless neutrino model and two massive neutrino scenarios 
(\({M_{\nu}=\qtylist{0.1;0.4}{\electronvolt}}\))---and analyze three redshift slices (\(z = \numlist{2;1;0}\)), 
averaging results over \num{100} realizations per cosmology at each redshift.

Among our key results (\Cref{sec:results-snapshots,sec:results-halos}), 
we anticipate that high-persistence apex positions in persistence diagrams 
for both the \gls{dm} and halo density fields will be particularly sensitive to neutrino masses, 
especially for saddle statistics at high redshift.
For the \gls{dm} density field, we find that Betti curves broaden and flatten with increasing neutrino mass, 
exhibiting two characteristic density thresholds where the Betti numbers remain invariant. 
Halo fields, despite their sparser sampling, also retain clear topological information, 
with direct implications for observational programs: we will show that Betti curves from halo fields 
exceed the fiducial model at \(z = 2\) up to \qty{10}{\percent} for \qty{0.1}{\electronvolt} neutrinos. 
Overall, the mass-dependent signatures identified from our persistent homology analysis reveal clear detectable 
neutrino effects in the cosmic web at the few-percent level, even for $M_\nu \sim 0.1$ eV---findings that we demonstrate for the first time in this work.

The results presented here will thus provide a solid foundation for forward-modeling or emulator-based approaches 
using persistent homology and environment-based metrics to constrain---or potentially detect---neutrino mass, alongside other cosmological parameters, 
with direct implications for ongoing and upcoming surveys (e.g., DESI, Euclid, Rubin-LSST). 
In particular, our study offers critical insights into the physical effects of massive neutrinos in the cosmic web 
and guides the interpretation of persistence-based constraints, representing a significant step toward robust, physically grounded cosmological measurements. 
Our results are also highly relevant for forthcoming observations of the high-redshift cosmic web, a relatively unexplored territory. Probing large-scale structure at 
$z \gtrsim 2$ is rapidly becoming feasible, and this regime is especially favorable for probing neutrinos because their associated signatures are 
cleaner and less affected by late-time nonlinearities, making it a very promising avenue for future investigations.

The paper is organized as follows. \Cref{sec:simulations} presents a brief overview of the simulations used in this study. 
\Cref{sec:methodology} outlines our methodology, key algorithms, and pipelines for constructing density fields, 
extracting persistent features, and computing Betti curves and other topological summaries. 
The main results---persistent diagrams, Betti curves, and additional topological statistics---are 
presented in \Cref{sec:results-snapshots} 
for \gls{dm} density fields derived from simulation snapshots, and in \Cref{sec:results-halos} 
for halo density fields obtained from halo catalogs. 
Finally, \Cref{sec:conclusions} summarizes our key findings and highlights ongoing and future work. 


\section{Simulations}     \label{sec:simulations}

In this study, we use a subset of the \Quijote{} simulations \citep{Quijote:2020} to compute topological statistics.
The \Quijote{} simulation suite is a set of large-volume \(N\)-body runs performed with the \gls{tpm} code \Gadget{}, 
a modified and improved version of \Gadget[II]{} \citep{Gadget-II}. Each simulation follows collisionless matter in a periodic comoving cube of 
side length \(L_{\mathrm{box}} = \qty{1}{\per\hubble\giga\parsec}\). The standard-resolution realizations contain \(N_{\mathrm{CDM}} = \num{512}^{3}\) \gls{cdm} 
particles. In models that include massive neutrinos, additional \(N_{\nu} = \num{512}^{3}\) neutrino particles are evolved as a separate, collisionless 
species whose initial thermal velocities are sampled from a Fermi-Dirac distribution. 
Gravitational forces are softened with a fixed 
Plummer-equivalent length set to \(\sfrac{1}{40}\) of the mean inter-particle separation.

The fiducial cosmology of the \Quijote{} simulations adopts parameters consistent with those reported by \citet{Planck:VI:2020}.
Specifically, the different parameters are: the total matter density \(\Omega_{\mathrm{m}} = \num{0.3175}\), the baryon density \(\Omega_{\mathrm{b}} = \num{0.049}\), 
the reduced Hubble constant \(h = \num{0.6711}\), the scalar spectrum power-law index \(n_{\mathrm{s}} = 0.9624\), the present root-mean-square 
matter fluctuation over \qty{8}{\per\hubble\mega\parsec} spheres \(\sigma_{8} = \num{0.834}\), and the DE equation-of-state (EoS) parameter \(w = \num{-1}\). 
The baseline model imposes a flat universe so that the \gls{de} density parameter is \(\Omega_{\Lambda} = 1 - \Omega_{\mathrm{m}} = \num{0.6825}\) (\cref{tab:quijote-simulations}). 
Furthermore, the fiducial cosmology does not include massive neutrinos.

All runs start from high redshift, \(z_{\mathrm{in}} = 127\). Transfer functions are computed with the Boltzmann solver \Camb{} \citep{CAMB}, 
and used to generate the initial particle displacements and velocities. For cosmologies without massive neutrinos, initial conditions 
are generated via \gls{2lpt} to reduce transient modes. For the massive neutrino runs, initial conditions are
instead generated with a \gls{za} \citep{Zel'dovich:1970}, supplemented by a rescaling procedure that consistently accounts for the scale-dependent 
linear growth introduced by neutrino free-streaming. Additionally, a matched set of fiducial \gls{za} realizations is included so that derivatives with respect to 
the sum of neutrino masses are not contaminated by differing transient behavior between models\@. In our analysis, we use these 
fiducial \gls{za} simulations for a consistent comparison with massive neutrinos simulations.

Notably, in the \Quijote{} suite, massive neutrinos are implemented as particles: 
discrete tracers that contribute to the gravitational potential and are assigned thermal velocities appropriate for 
\(z_{\mathrm{in}} = 127\). 
Their masses are assumed degenerate across the three active species, so only the total mass, \(M_{\nu}\equiv\sum_{i} m_{\nu,i}\), is considered.

In this work, we focus on three representative runs: the fiducial or \textit{baseline}, Planck-like, massless neutrino cosmology, and two massive neutrino 
scenarios with \({M_{\nu}=\qtylist{0.1;0.4}{\electronvolt}}\), respectively. All realizations analyzed here use the standard-resolution setting described above. 
We select simulations that adopt the same initial conditions method and random seeds across different cosmologies, ensuring that morphological 
differences are driven by changes in neutrino masses rather than sample variance or transient effects. 
\Cref{tab:quijote-simulations} summarizes the main parameters and properties of the simulation subset.

The outputs of the \Quijote{} realizations include snapshots containing full particle data, as well as 
halo catalogs extracted with \Gadget{} using the \gls{fof} method.
From the \num{500} runs in each set, we select a subset of \num{100}  with indices in the range \numrange{100}{199}. 
Our analysis focuses on three redshifts: \(z = \num{2}\), \(\num{1}\), and \(\num{0}\).
Details of the post-processing pipeline applied to these simulations are presented next.

\begin{table}[tp!]
\caption{Summary of the \Quijote{} simulations used in our analysis. 
The names in the \textit{``Varying Parameters''} section correspond to those in 
Table 1 of \citet{Quijote:2020}. 
The \textit{``Realizations''} row gives the indices of the runs selected 
from the 500 simulations per cosmology in the suite.} 
\label{tab:quijote-simulations}
\centering
\begin{tabular}{lc}
\bottomrule
\multicolumn{2}{c}{Fixed Parameters}                             \\
\midrule
\(\Omega_{\mathrm{m}}\) & \num{0.3175}                           \\
\(\Omega_{\mathrm{b}}\) & \num{0.049}                            \\
\(\Omega_{\Lambda}\)    & \num{0.6825}                           \\
\(h\)                   & \num{0.6711}                           \\
\(n_{\mathrm{s}}\)      & \num{0.9624}                           \\
\(w\)                   & \num{-1}                               \\
\bottomrule
\multicolumn{2}{c}{Varying Parameters}                           \\
\midrule
Fid                     & \(M_{\nu} = \qty{0.0}{\electronvolt}\) \\
\(M_{\nu}^{+}\)         & \(M_{\nu} = \qty{0.1}{\electronvolt}\) \\
\(M_{\nu}^{+++}\)       & \(M_{\nu} = \qty{0.4}{\electronvolt}\) \\
\bottomrule
\multicolumn{2}{c}{Simulation Details}                           \\
\midrule
\(L_{\mathrm{box}}\)    & \qty{1}{\per\hubble\giga\parsec}       \\
\(N_{\mathrm{CDM}}\)    & \(\num{512}^{3}\)                      \\
\(N_{\nu}\)             & \(\num{512}^{3}\)                      \\
\Glsxtrshortpl{ic}      & \Glsxtrshort{za}                       \\
\(z_{\mathrm{in}}\)    & \num{127}                              \\
Realizations            & \numrange{100}{199}                    \\
 \bottomrule
\end{tabular}
\end{table}


\section{Methodology}     \label{sec:methodology}

In this section, we briefly summarize our methodology, key algorithms, and pipelines for 
constructing density fields, extracting 
persistent features, and computing Betti curves and other topological summaries.
The primary approach adopted in this work, extensively described in \citet{Rossi:2026}, is based on discrete 
Morse theory and employs a scale-adaptive, parameter-free formalism.  
It involves computing the discrete Morse-Smale complex, combined with topological simplification via persistence. 
In addition, we also use a complementary technique, referred to as the \textit{``density-threshold-based''} 
approach in configuration space, as presented in a related study \citep{Moon:2023}. 
For a detailed description of these methods, we refer the reader to \citet{Rossi:2026} and \citet{Moon:2023}.

\subsection{Construction of Density Fields}       \label{subsec:constructing-density-fields}

Extracting topological information requires a density field; therefore, we first explain how 
these fields are constructed from simulation snapshots or halo catalogs. 
We use \DisPerSE{} \citep{DisPerSE:2011b} as our baseline algorithm for characterizing the discrete Morse-Smale complex. 
\DisPerSE{}'s \gls{mse} operates on a simplicial complex representation of the density field. 
One way to construct such a field is by computing a triangulation with particles as vertices and estimating 
the density at each vertex based on the volumes of the surrounding tetrahedra. 
However, given the large number of particles in each \Quijote{} snapshot, 
running \DisPerSE{} on full snapshot triangulations is challenging due to computational memory limits.  
Using a random subsample of particles could alleviate the memory issue, but at the cost of added noise.
Instead of discarding particles, we compute the density field of each simulation with a particle-mesh-based method to 
reduce the number of vertices, following the approach of \citet{Moon:2023}.
The resulting density field is a regular, cubic three-dimensional (3D) grid of vertices that \DisPerSE{} can implicitly triangulate.
Details of our procedure for snapshot density estimation are provided in \cref{subsubsec:density-from-snapshots}.
In contrast, halo catalogs contain far fewer particles, allowing direct tessellation with a \gls{dtfe}, as described in \cref{subsubsec:density-from-halos}, 
following the approach of \citet{Rossi:2026}. 

\subsubsection{From Simulation Snapshots}          \label{subsubsec:density-from-snapshots}

\begin{figure*}[htp!]
\centering
\includegraphics[width=\textwidth]{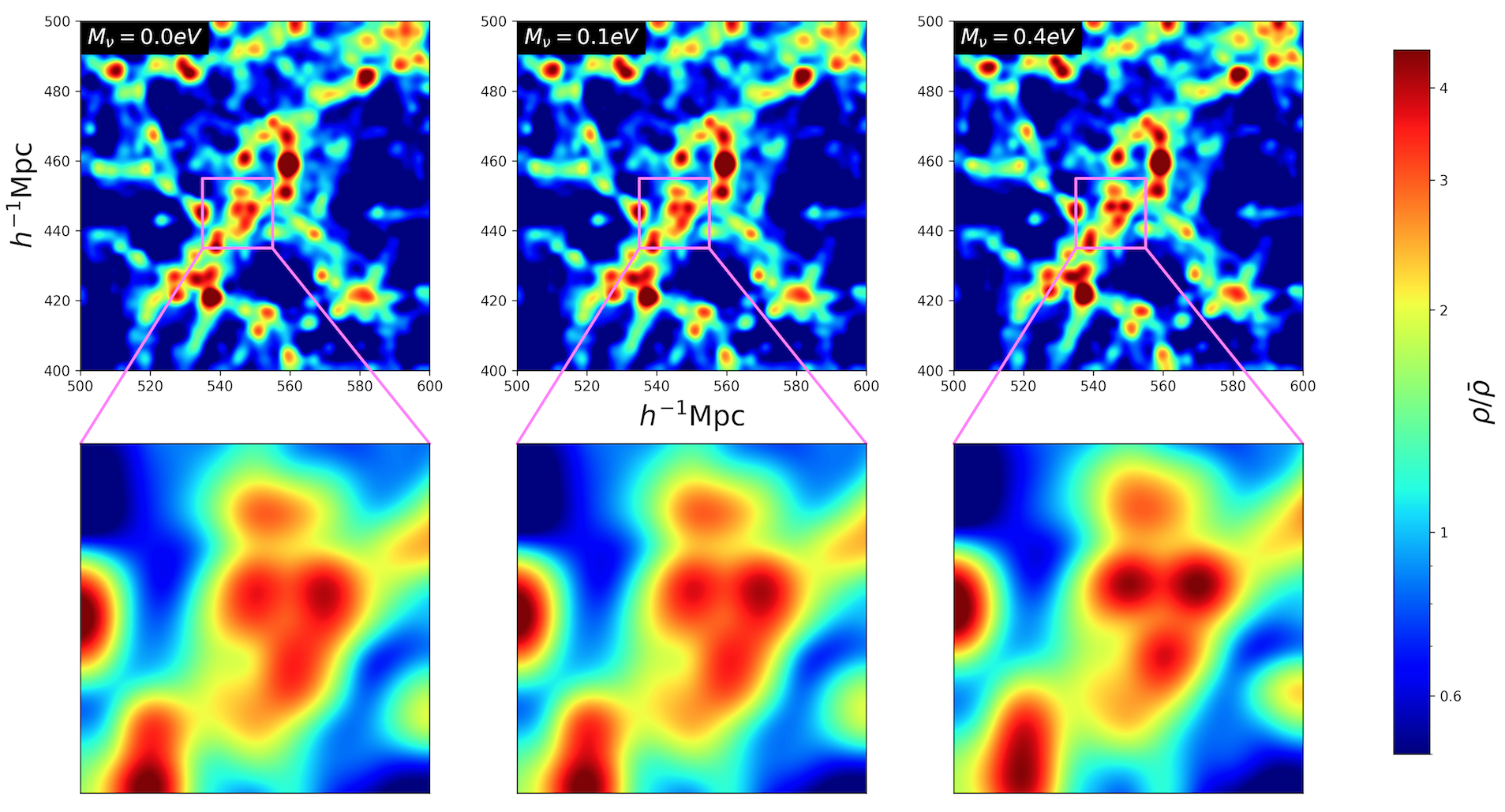}
\caption{Projection of the \gls{dm}  
density field at \(z = 0\) 
from \Quijote{} simulations with different neutrino masses. Panels, from left to right, correspond to 
\(M_{\nu} = \qtylist{0.0;0.1;0.4}{\electronvolt}\), respectively. 
Top panels show a \qtyproduct{100x100}{\per\hubble\mega\parsec} area, 
averaged over a \qty{50}{\per\hubble\mega\parsec} thick slice.
Bottom panels show corresponding
zoom-ins of the \qtyproduct{20x20}{\per\hubble\mega\parsec} region indicated by squares on the top panels.
Morphological differences are clearly visible across the different cosmological models.}
\label{fig:quijote-projection}
\end{figure*}

We use the density construction routine from \Pylians{} \citep{Pylians:2018}, which is 
based on particle-mesh mass-assignment methods. 
\Pylians{} is a set of Python libraries designed for the analysis of  \(N\)-body  and hydrodynamical numerical simulations.
We employ a \gls{cic} mass-assignment scheme on a cubic grid of  \(128^{3}\) cells, 
corresponding to an average of \num{64} \gls{dm}  
particles per cell and a spatial resolution of \qty{7.8125}{\per\hubble\mega\parsec}. 
In this scheme, each particle is assumed to occupy the volume of a single grid cell, with its mass evenly distributed and centered on the particle position. 
Fractions of the particle mass are assigned to the eight surrounding cells according to the overlap between the particle's shape and the cell volumes.
A similar density construction procedure was adopted in \citet{Moon:2023}, although in that case a triangular-shape cloud (TSC) 
mass-assignment scheme was instead used.

\cref{fig:quijote-projection} shows a visualization of such a density field, from three selected  \Quijote{} snapshots at \(z = 0\). The top row displays the projection of 
a \qty{50}{\per\hubble\mega\parsec}-thick slice over a \qtyproduct{100x100}{\per\hubble\mega\parsec} area for one simulation from each set: 
massless neutrinos on the left, massive \qty{0.1}{\electronvolt} neutrinos in the middle, and \qty{0.4}{\electronvolt} neutrinos on the right. 
The three simulations share the same initial random noise, allowing similar large-scale structures to be seen in each model, with small differences 
arising from the varying neutrino masses. The bottom row shows zoom-ins of the regions highlighted by squares in the top panels, 
where subtle differences in shape and density contrast are clearly visible.

\subsubsection{From Halo Catalogs}       \label{subsubsec:density-from-halos}

The number of halos in each \Quijote{} catalog is much smaller than the number of \gls{cdm} 
and neutrino particles in the corresponding snapshot. 
There are approximately \num{40000} halos at \(z = 2\), and about \num{400000} at \(z = 0\). 
This reduced number of points, together with the fact that each halo has a different size, makes the 
particle-mesh method used for snapshot particles much less suitable. 
The shot noise would be higher, larger cells would be required, and some topological information would be most certainly lost, as each halo would 
effectively be assigned the same shape determined by the mass-assignment scheme and grid size. Hence, 
halo catalogs are better suited for a full \gls{dtfe} treatment,
following the methodology described in \citet{Rossi:2026}. 

We therefore employ the \gls{dtfe} approach to estimate the density field from halo catalogs. The \gls{dtfe} is a self-adaptive technique 
that reconstructs a piecewise linear density field from a set of point positions and masses, naturally capturing the anisotropic and 
hierarchical properties of cosmic structure. The process begins with a Delaunay tessellation of the particle positions, decomposing the 
simulation domain into non-overlapping tetrahedra with the particles as vertices. 
The local density at each particle is then estimated as the inverse of the volume of 
its adjacent tetrahedra, multiplied by the particle mass.

We use the \texttt{delaunay\_3D} tool provided by \DisPerSE{}. First, we convert the \Quijote{} FOF
halo catalogs into \DisPerSE{} \texttt{NDfield} files 
containing the halo positions. We then run the \texttt{delaunay\_3D} tool to compute the 
Delaunay triangulation and estimate the density at each node. 
The output is a \DisPerSE{} \texttt{NDnet} file containing the list of nodes (halo positions) 
and the associated edges, triangles, and tetrahedra. 
The file also includes a node-associated field 
representing the number 
density, i.e., the inverse volume of the tetrahedra surrounding each node. 
We scale this field by the corresponding halo mass at each node to obtain a mass density estimate, 
which is then used for persistent feature extraction.

\subsection{Extraction of Critical Points and Persistent Features}        \label{subsec:extracting-features}

Once a density field is constructed, either on a regular grid or as a Delaunay tessellation, we use \DisPerSE{}'s \texttt{mse} 
tool to extract topological information. This tool computes the Morse-Smale complex, 
which partitions space based on ascending and descending manifolds. 
In addition, \DisPerSE{} identifies persistent pairs of critical points.
In three dimensions, there are four types of critical points, each characterized by an associated order \(k\). 
A minimum has order 0, a maximum has order 3, and saddle points have order 1 or 2. 
In the context of the cosmic web, a maximum corresponds to a matter overdensity or peak (e.g., a dark matter halo). 
A saddle-2 is associated with a filament connecting different maxima, while a saddle-1 corresponds to a wall 
linking filaments and surrounding cosmic voids. 
Voids are associated with minima. Each pair of critical points consists of two points whose orders differ by 
one and is assigned a persistence value that quantifies its significance. 
Small-scale fluctuations produce pairs with low persistence, whereas the most prominent large-scale topological 
features correspond to pairs with high persistence.

Following \citet{DisPerSE:2011b}, our persistence diagrams (e.g., bottom panels of \cref{fig:persistence-betti-construction}) 
represent persistence pairs 
by plotting the lower density of each pair on the horizontal axis and the higher density on the vertical axis. 
We interpret the lower density as the feature \textit{birth} density and the higher density as the feature \textit{death} density 
(see \cref{subsec:computing-betti-curves} for more details). With this convention, we are effectively analyzing 
the topology of the \textit{sublevel set}. Compared to the \textit{superlevel set}, which is its complement, the Betti curves have similar shapes, 
but with some labels exchanged.

For periodic boundary conditions, as in our case, the topologies of the sublevel and superlevel sets are perfectly symmetric: 
a hole in one corresponds to an island in the other, a wall to a filament, and vice versa. 
For example, \citet{Wilding2021} adopt the superlevel set convention in their study, which results in 
a swap of the birth and death axes in persistence diagrams and a corresponding relabeling 
of the Betti curves.\footnote{What \citet{Wilding2021} denote as 
\({\mathtt{dim} = \num{0}}\), \(\num{1}\), 
and \(\num{2}\), correspond in our notation to  
\(P_{2}\), \(P_{1}\), and \(P_{0}\), respectively.} 

\subsection{Betti Curve Computation}         \label{subsec:computing-betti-curves}

\begin{figure}[htp!]
\centering
\includegraphics[width=\columnwidth]{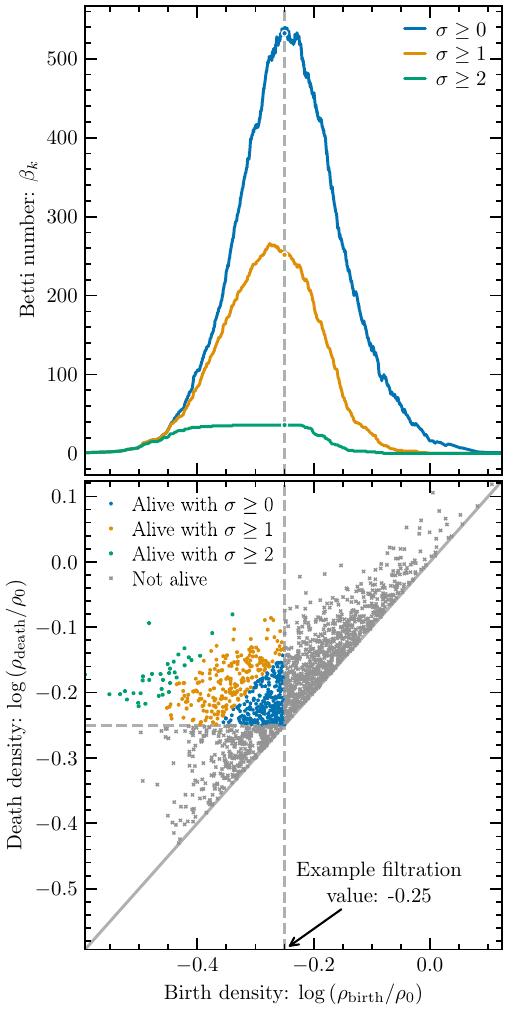}
\caption{Illustration of the construction of Betti curves (top panel) 
from the persistence diagram (bottom panel) for a given pair of type \(k\).
The vertical dashed line indicates an example filtration value. 
The colors and shapes of points in the diagram show which pairs are counted or not at this 
filtration value for different persistence
(\(\sigma\)) levels. See the main text in \cref{subsec:computing-betti-curves} for details.}
\label{fig:persistence-betti-construction}
\end{figure}

Betti numbers are fundamental topological invariants that quantify the number of independent, 
nontrivial holes or cycles in different dimensions within a topological space. 
They are essential for characterizing the shape and connectivity of complex structures, such as those present in cosmological density fields. 
In a 3D field, there are three non-zero Betti numbers: the zeroth-dimensional Betti number \(P_{0}\) 
counts the number of connected components,
the first-dimensional number \(P_{1}\)
counts one-dimensional loops, 
and the second-dimensional number \(P_{2}\) 
counts cavities.

To compute the persistence diagrams and Betti curves, we use a filtration based on the density field sublevel set. 
As the filtration value increases, when it reaches a local minimum of the density field, a 
connected component of the sublevel set appears, increasing 
\(P_{0}\) by one.
Two connected components merge at a saddle-1 critical point (i.e., a wall), decreasing 
\(P_{0}\) by one. This merging can create a loop, increasing 
\(P_{1}\) by one.
A loop can then collapse at a saddle-2 critical point (i.e., a filament), decreasing 
\(P_{1}\) by one. 
The collapse of a loop can form a cavity, increasing 
\(P_{2}\) by one, which subsequently disappears when the filtration reaches a maximum, decreasing 
\(P_{2}\) by one.

We calculate the Betti numbers for a given density threshold using the list of critical point pairs produced by \DisPerSE{}. 
Each pair connects a critical point of order \(k\)  to one of order  \(k + 1\): minimum to saddle-1 (\(P_{0}\)),
saddle-1 to saddle-2 (\(P_{1}\)), 
and saddle-2 to maximum (\(P_{2}\)).
In each pair, the lower- and higher-order critical points correspond, respectively, to the appearance 
and disappearance---or {\itshape birth} and {\itshape death}---of a topological feature. 
For a given filtration value, we count the number of pairs of each type that are {\itshape alive} at that threshold. 
The full Betti curves then represent the Betti numbers as a function of the filtration parameter, which in our case is the density.

\cref{fig:persistence-betti-construction} shows an example of a persistence diagram (bottom panel) and the corresponding Betti curves (top panel), 
with annotations illustrating the pair-counting procedure. The diagram can represent any of the three pair types, and the process is repeated 
separately for each to produce three distinct Betti curves. The persistence diagram on the bottom is a scatter 
plot in which each point corresponds to a persistence pair. 
Both axes represent density: the horizontal axis shows the birth density of a pair, and the vertical axis shows its death density. 
Since the death density is always greater than the birth density, all points lie in the upper-left triangle of the diagram.
We illustrate one example filtration value with a vertical dashed line. At this threshold, any pair to the right of the line 
has not yet been born and is shown as a gray cross. Pairs to the left of the line are already born; 
however, those below the horizontal dashed line have already died and are also shown as gray crosses. 
The remaining pairs are alive at the given filtration value, and their count defines the Betti number. 
Counting all such living pairs produces one point on the blue Betti curve shown 
in the top panel of \cref{fig:persistence-betti-construction}. 
 
Each pair has an associated  ``number of \(\sigma\)'' 
that quantifies its significance.
Pairs with higher \(\sigma\) 
values are more likely to correspond to prominent topological features rather than noise. 
In \cref{fig:persistence-betti-construction}, living pairs with 
\(\sigma\) > \num{2} are shown in green, those with 
\(\sigma\) > \num{1} in orange, and the remaining pairs in blue. 
The simplification process consists of removing persistent pairs with 
\(\sigma\) below a chosen threshold. 
In the figure, the blue Betti curve is obtained by counting all living pairs. The orange curve includes only pairs with 
\(\sigma\)  > \num{1} (orange and green points), 
and the green curve includes only pairs with 
\(\sigma\) > \num{2}  (green points).

\begin{figure}[tp!]
\centering
\includegraphics[width=\columnwidth]{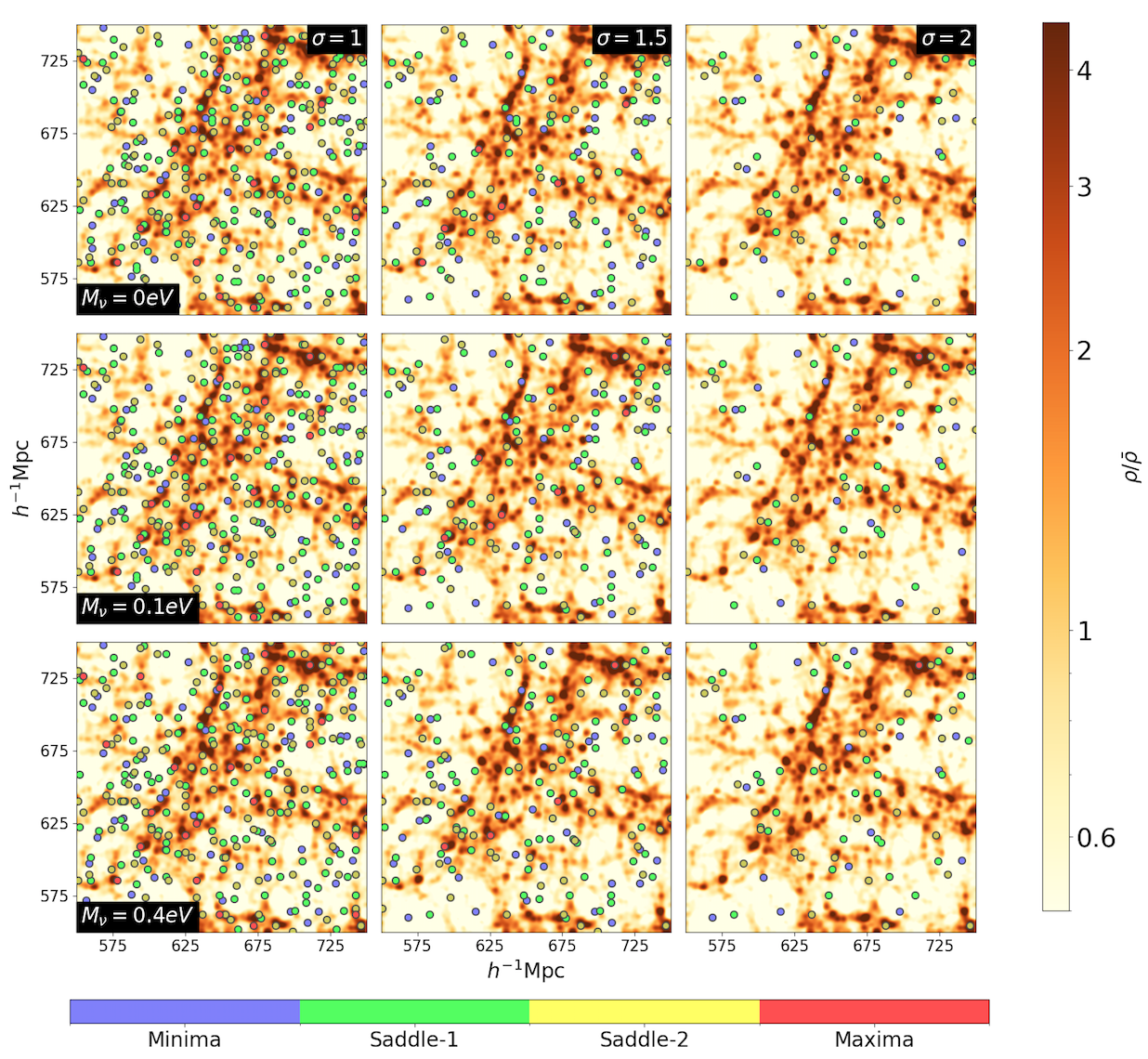}

\caption{Projection of  \qtyproduct{200 x 200}{\per\hubble\mega\parsec}, \qty{30}{\per\hubble\mega\parsec} thick slices 
of \Quijote{}  \gls{dm}   
density fields, sharing the same initial random noise, with overlaid critical point positions at \(z = \num{0}\). 
The top row shows a fiducial, massless neutrinos simulation, while the middle and bottom rows correspond to 
\qty{0.1}{\electronvolt} and \qty{0.4}{\electronvolt} massive neutrino simulations, respectively---similarly to \cref{fig:quijote-projection}. 
Each column applies a different simplification  \(\sigma\) 
threshold: \num{1} on the left, \num{1.5} in the middle, and \num{2} on the right. 
As the threshold increases, the number of critical points decreases, leaving only those belonging to the most persistent pairs.}
\label{fig:quijote-critical-points}
\end{figure}

The simplification process is also illustrated in \cref{fig:quijote-critical-points}, which shows slice projections of 
 \Quijote{}  \gls{dm}  
 density fields---sharing the same initial random noise---at \(z = \num{0}\) with the positions of critical points overlaid. 
Each row corresponds to a different cosmological model: fiducial, massless neutrinos on top; 
\qty{0.1}{\electronvolt} massive neutrinos in the middle; and \qty{0.4}{\electronvolt} massive neutrinos on the bottom (as in \cref{fig:quijote-projection}). 
Each column corresponds to a different simplification \(\sigma\) 
threshold: \num{1} on the left, \num{1.5} in the middle, and \num{2} on the right. 
Only critical points in pairs with  \(\sigma\) values exceeding the given threshold are shown. 
At lower  \(\sigma\) values, more critical points remain, but these include less significant structures and noise. 
As the \(\sigma\) threshold increases, weaker points are gradually removed, leaving only the most prominent features. 
Additionally, as the neutrino mass increases, the overall number of critical points decreases, particularly for small-scale structures, 
a trend demonstrated in detail by \citet{Moon:2023}.


\section{Results from the Dark Matter Density Field}        \label{sec:results-snapshots}     

This section presents the results obtained from \gls{dm}    
density fields computed from simulation snapshots, 
employing the methodology detailed in \cref{subsubsec:density-from-snapshots}.

\subsection{Persistence Diagrams}    \label{subsec:results-snapshots-persistence-diagrams}

\begin{figure}[htp!]
\centering
\includegraphics[trim=16 50 9 55, clip, width=0.90\columnwidth]{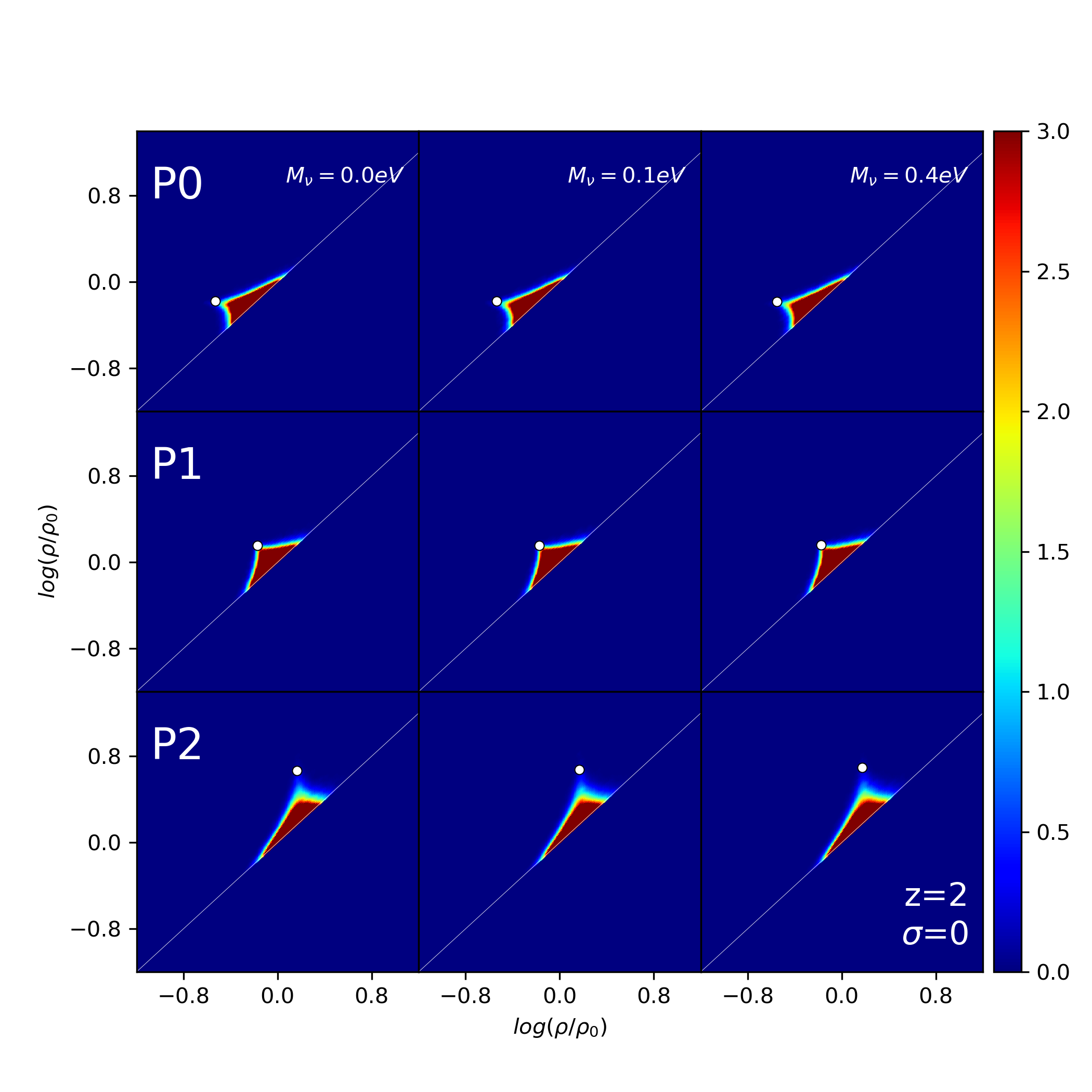}  
\includegraphics[trim=16 50 9 55, clip, width=0.90\columnwidth]{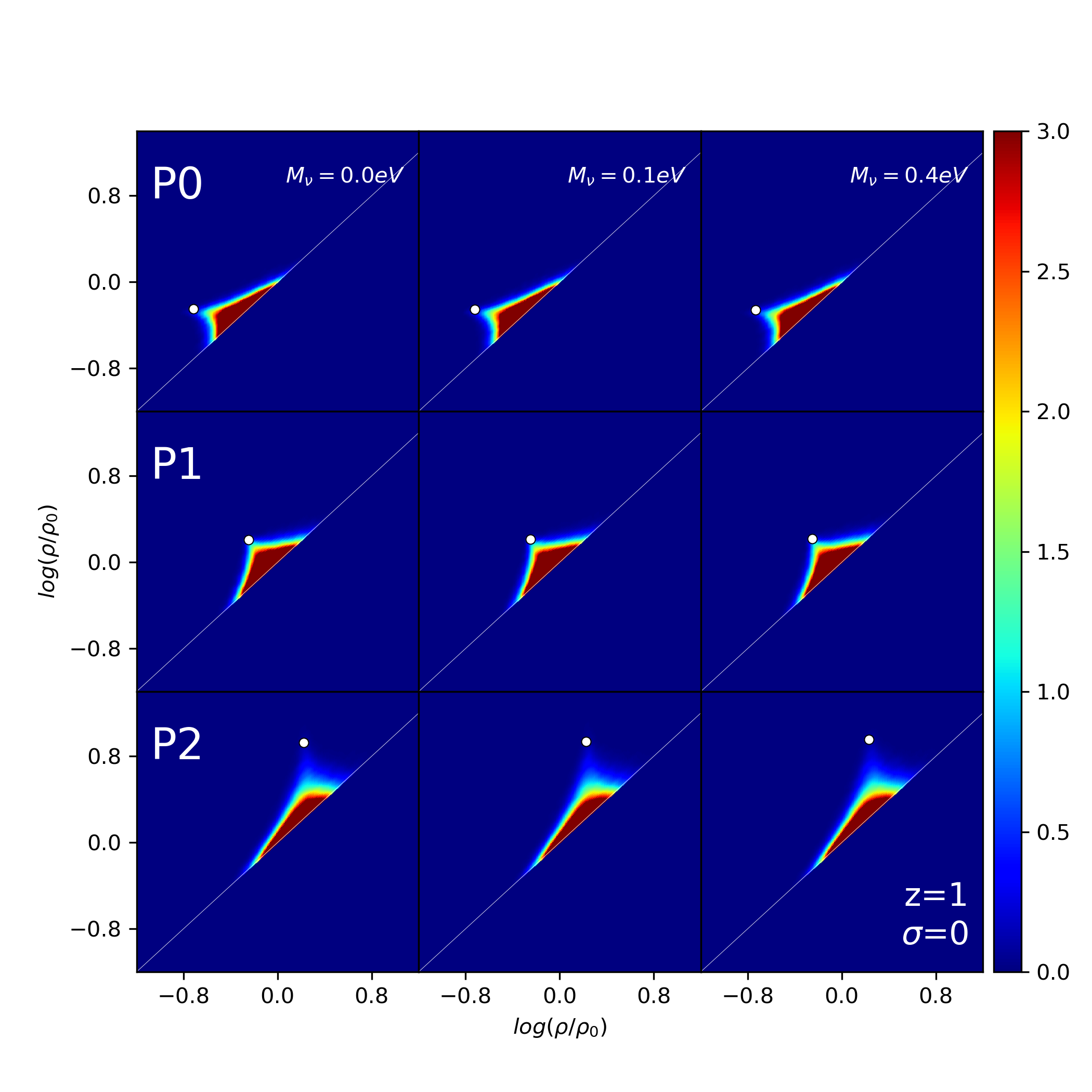} 
\includegraphics[trim=16 23 9 55, clip, width=0.90\columnwidth]{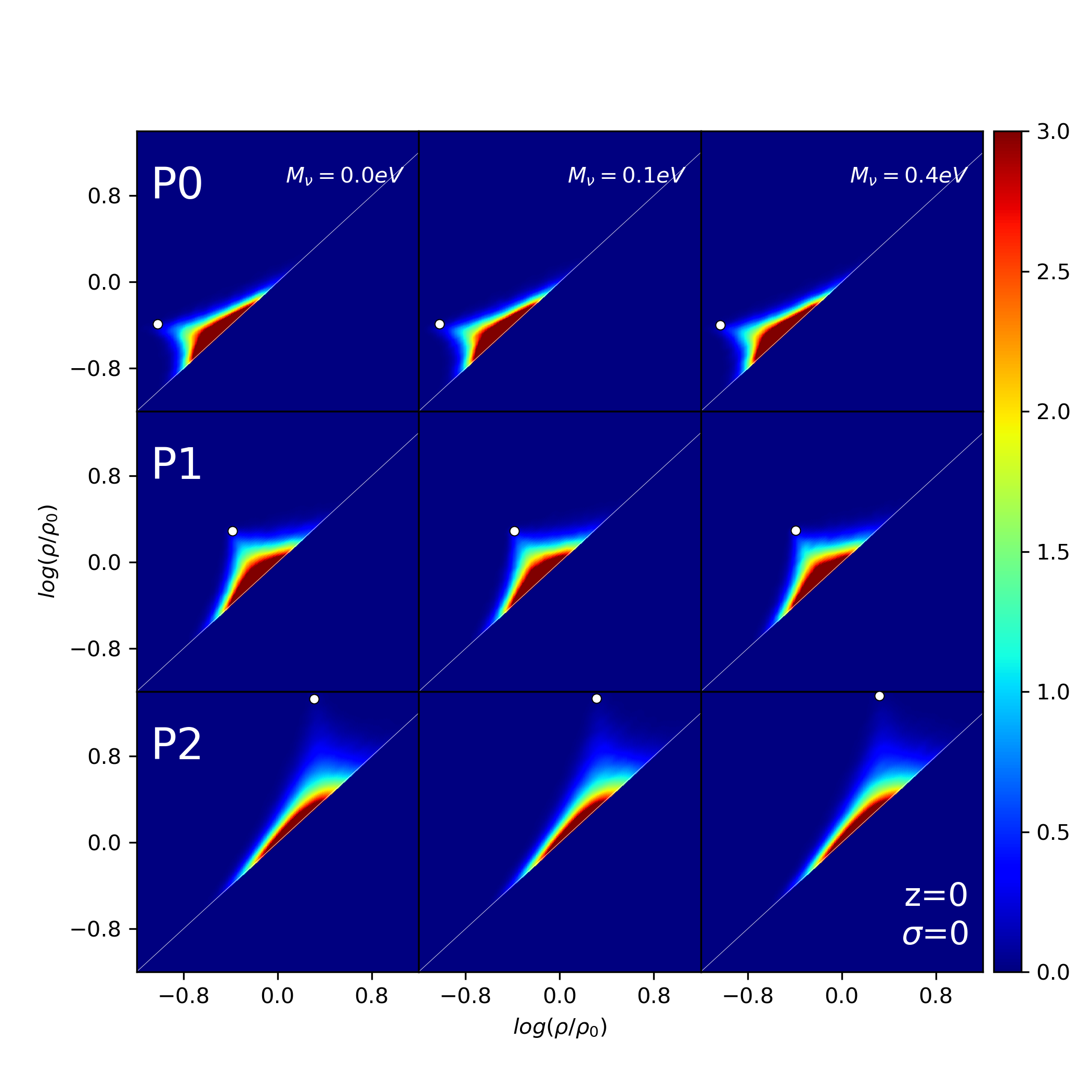}
\caption{Mean persistence diagrams for \Quijote{} \gls{dm}  
density fields at 
\(z = \num{2}\) (top block), \num{1} (middle block), and \num{0} (bottom block), 
averaged over \num{100} simulations per model. Colors indicate the number density of persistence pairs. 
No simplification is applied (\(\sigma\)=\num{0}). 
Each row shows a different pair type, and each column corresponds to a different cosmology, 
as specified in the panels. See the main text for additional details.}
\label{fig:particles-persistence}
\end{figure}  

We begin by examining the persistence diagrams derived from \Quijote{} \gls{dm}  
density fields, which summarize their key topological features. 
As described in \cref{sec:simulations}, our analysis considers two massive neutrino scenarios, 
\({M_{\nu}=\qtylist{0.1;0.4}{\electronvolt}}\), which we compare to a baseline Planck-like cosmology with massless neutrinos, 
using \num{100} realizations per model.

In \Cref{fig:particles-persistence}, we present the mean persistence diagrams for each type of persistence pair (rows) and 
for the two different neutrino mass models along with the fiducial case (columns) at three redshift intervals: 
\(z = \num{2}\) (top block), \num{1} (middle block), and \num{0} (bottom block).
In these diagrams, each point represents a persistence pair, 
with its distance from the diagonal reflecting its significance: points closest to the diagonal correspond to the least persistent features 
and typically trace noise. Given the large number of pairs, we compute 
two-dimensional histograms of the pair number density for each snapshot and then average them over \num{100} realizations
for a given cosmology. Following our convention described in \cref{subsec:extracting-features}, the birth density at the tip of the 
\(P_{0}\) diagram characterizes the typical void density, while the death density corresponds to matter walls. For 
\(P_{1}\), the tip's birth and death densities correspond to walls and filaments, respectively. For \(P_{2}\), 
the tip's birth and death densities correspond to filaments and matter density peaks.	 

The diagrams exhibit a clear triangular shape across all redshifts, with their bases along the low-persistence line---as noted by \citet{Wilding2021}. 
\(P_{0}\)  diagrams are skewed toward lower birth densities, 
\(P_{2}\)  toward higher birth densities, and 
\(P_{1}\)  diagrams are centered. This reflects the hierarchy of structure formation: 
\(P_{1}\)  pairs appear once 
\(P_{0}\)  pairs start disappearing, and 
\(P_{2}\)  pairs appear after   \(P_{1}\) 
pairs begin to vanish. 
The triangular shape traces phase transitions in the sublevel set at different density thresholds: for example, 
\(P_{1}\) pairs appear at one threshold, forming one side of the triangle, and disappear at another, forming the other side. 
The triangle's tip thus corresponds to characteristic density thresholds where these transitions occur.

\begin{figure}[htp!]
\centering
\includegraphics[trim=14 87 50 60, clip, width=\columnwidth]{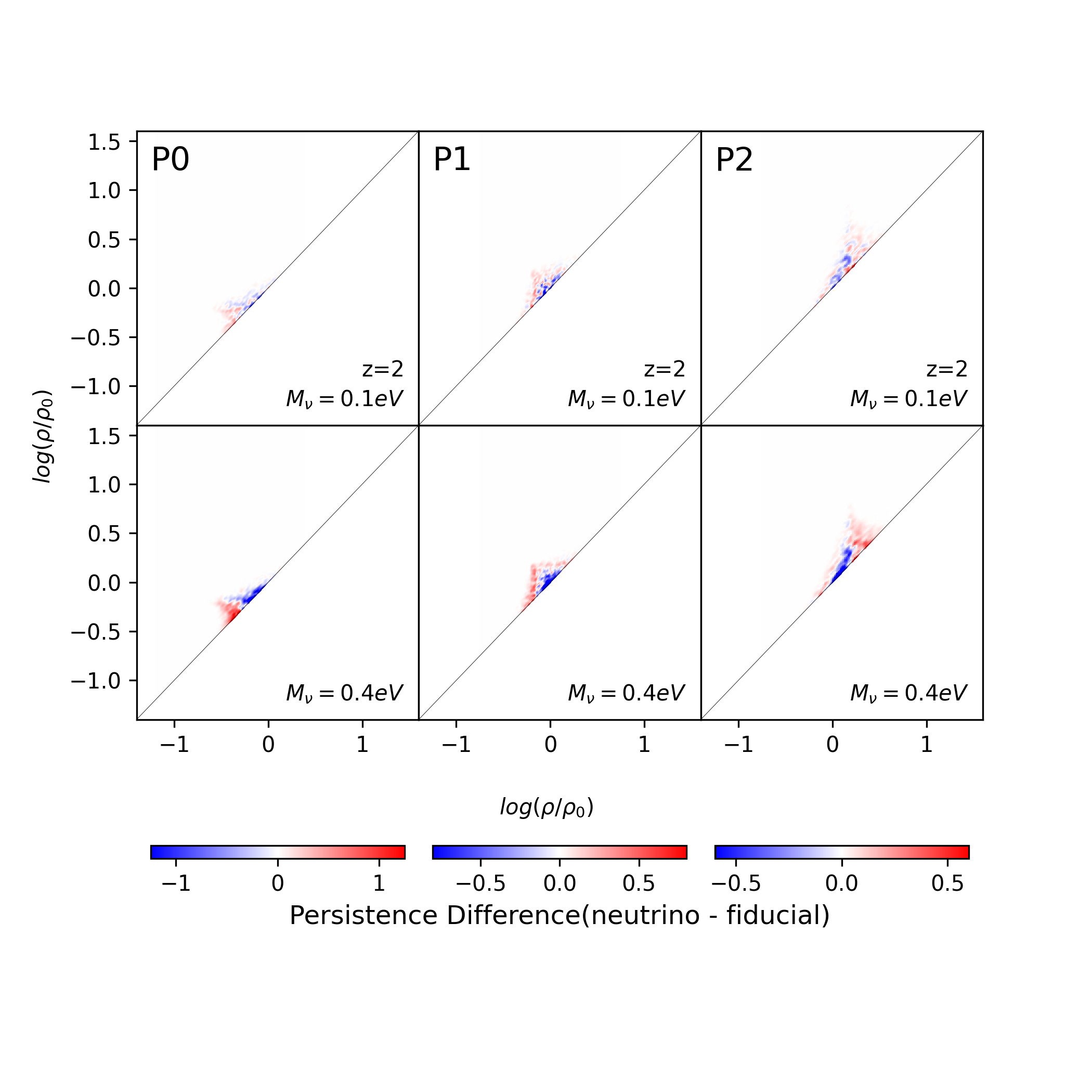}
\includegraphics[trim=14 87 50 60, clip, width=\columnwidth]{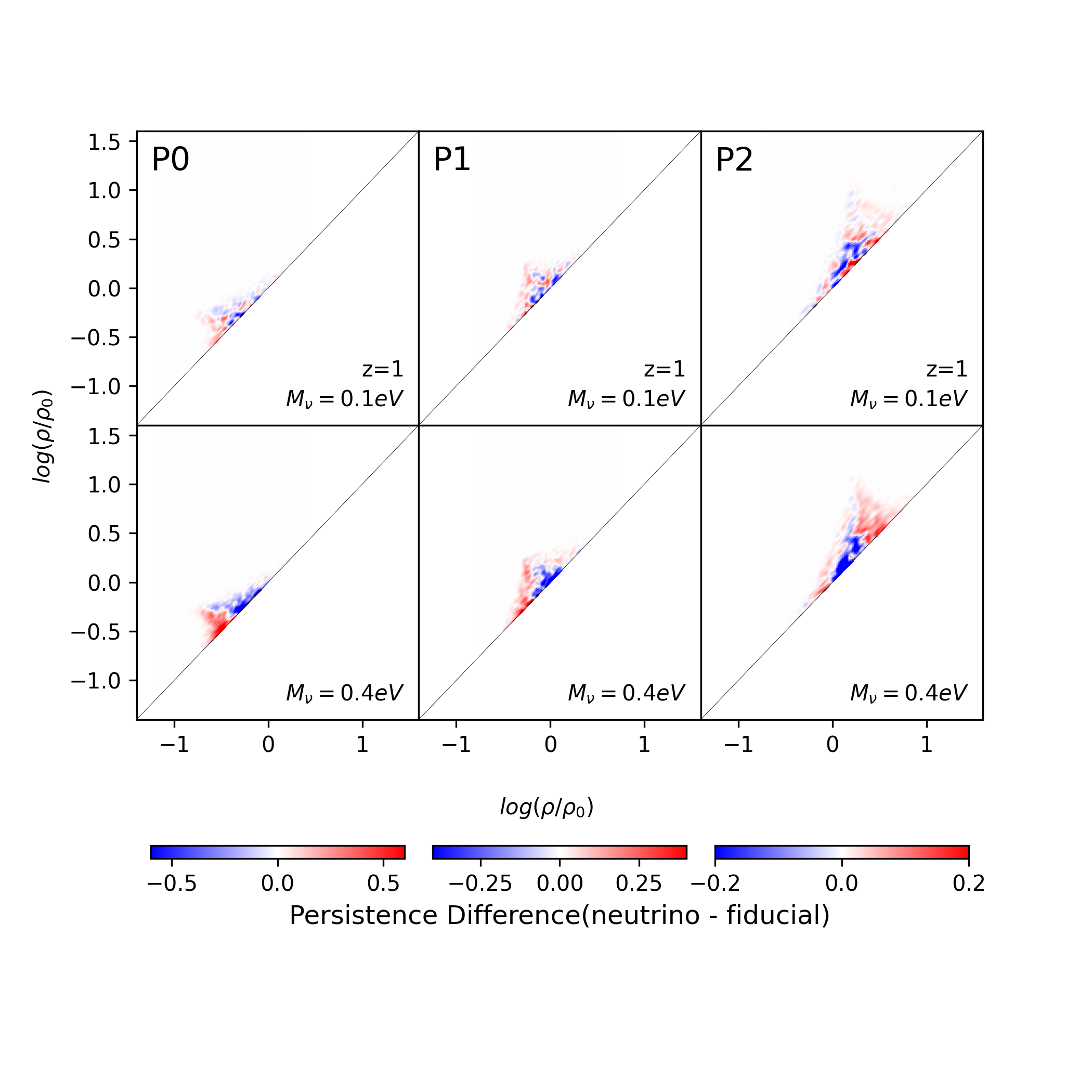}
\includegraphics[trim=14 77 50 60, clip, width=\columnwidth]{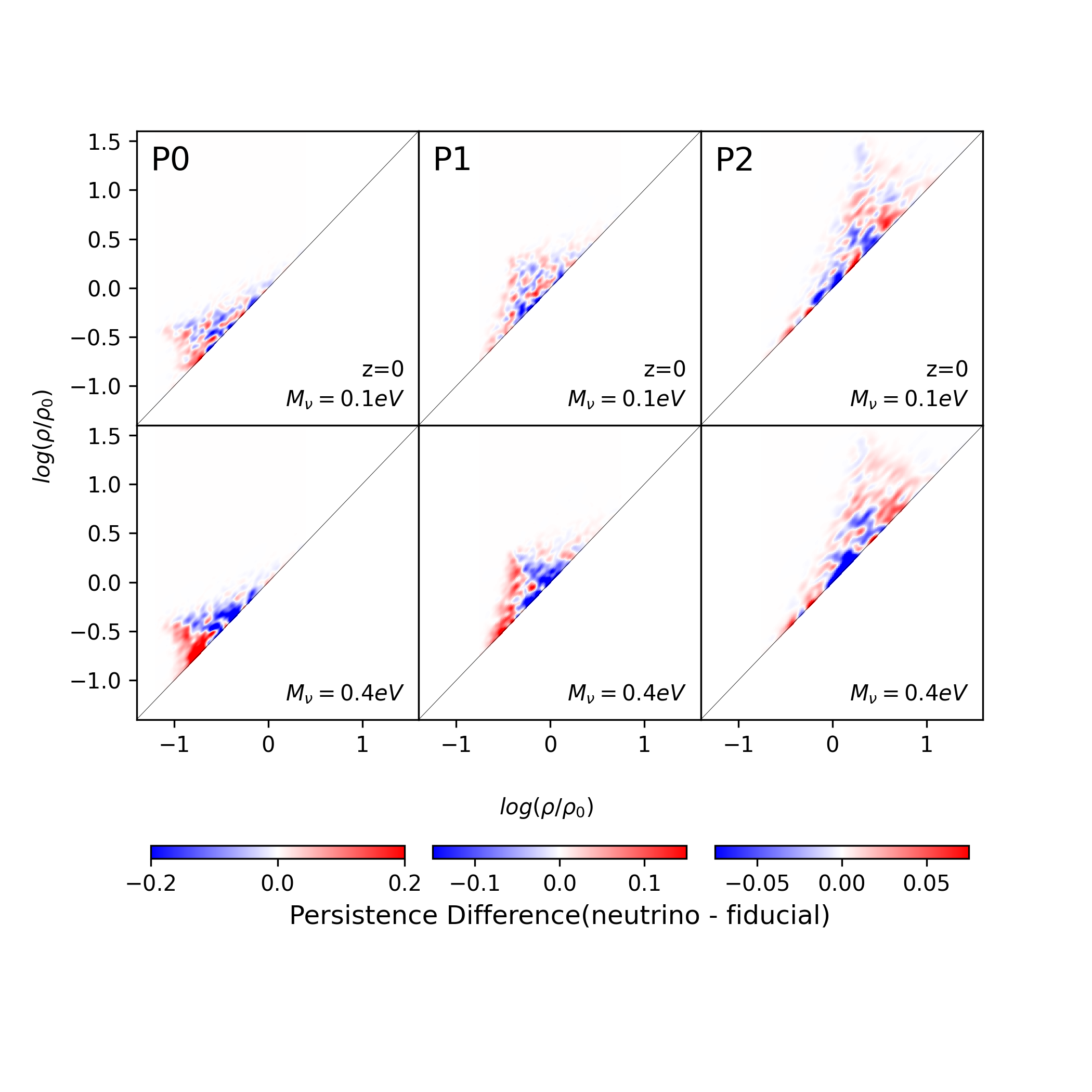}
\caption{Difference in the number density of persistence pairs between each massive neutrino model 
(\({M_{\nu}=\qtylist{0.1;0.4}{\electronvolt}}\)) and the fiducial, massless neutrino scenario in the birth-death plane 
at \(z = \num{2}\) (top block), \num{1} (middle block), and \num{0} (bottom block), for \Quijote{} \gls{dm}  
density fields. Colors indicate the pair number density difference, computed using a Gaussian kernel density estimator (KDE).}
\label{fig:particles-persistence-difference}
\end{figure}

In \cref{fig:particles-persistence}, differences between cosmologies are not easily discernible. 
To highlight them, \cref{fig:particles-persistence-difference} shows the same persistence diagrams as two-dimensional histograms, 
where each bin is colored according to the difference in persistence pair number density between massive neutrino models and the fiducial, 
massless neutrino case. In this figure, each column corresponds to a different pair type, and each row to a different neutrino mass.
For \(P_{0}\) (void-wall pairs), we observe an excess of pairs at low birth densities in the massive neutrino simulations, 
accompanied by a deficit at high birth densities; this effect becomes more pronounced with increasing neutrino mass. 
For \(P_{1}\),  the number density increases along a vertical band at low birth densities and a horizontal band at higher death densities. 
For \(P_{2}\)  (filament-peak pairs), a similar trend to \(P_{0}\) is observed, but with the excess shifted toward higher birth densities.
These features translate into the Betti curves as higher Betti numbers at low and high densities, and lower values at intermediate densities for massive neutrino models. 
This behavior is indeed confirmed in \cref{fig:particles-betti} (see the details in \cref{subsec:results-snapshots-betti-curves}).

\begin{figure*}[htp!]
\centering
\includegraphics[angle=0,width=0.93\textwidth]{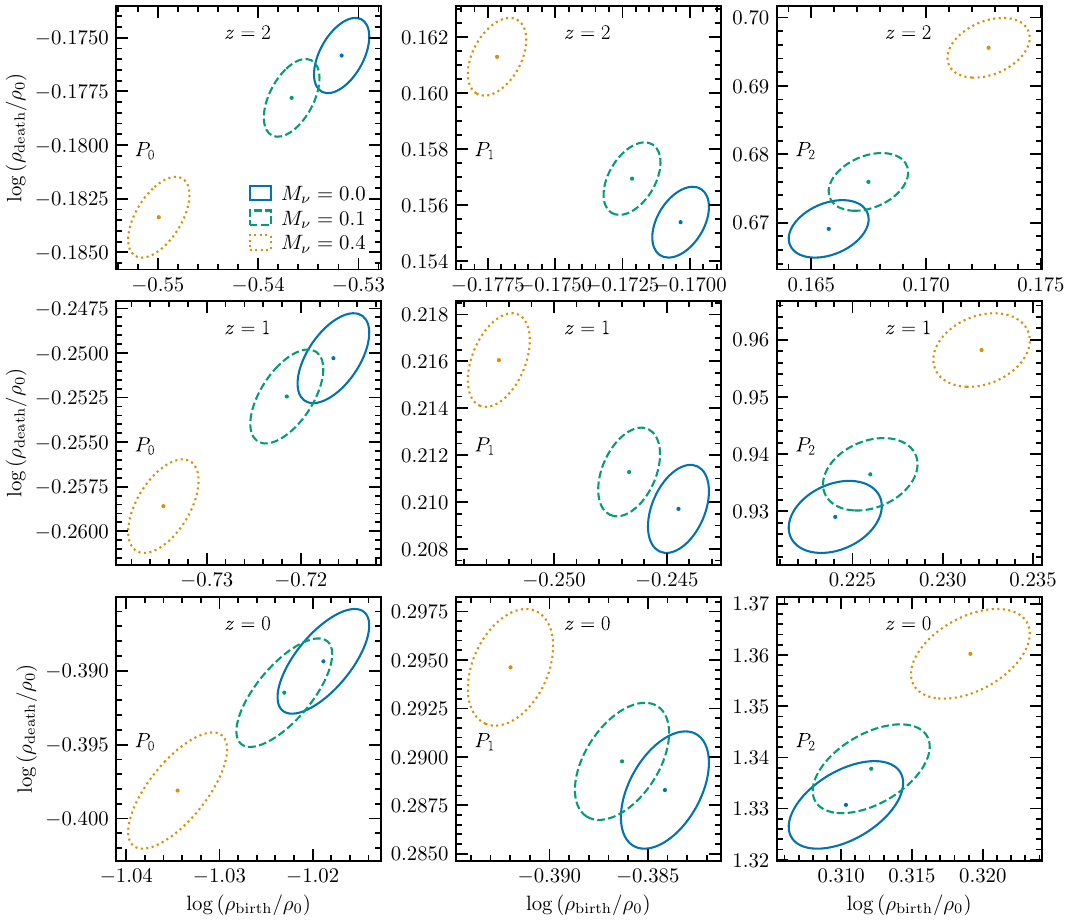}
\caption{Mean positions (points) of the persistence diagram apexes and their \num{1}\(\sigma\) 
variations (ellipses) over \num{100} simulations per cosmology for the \Quijote{} \gls{dm}  
density fields. Panels are arranged from top to bottom by redshift 
(\(z = \numlist{2;1;-0}\)) and from left to right by pair type (\(P_{0}\), \(P_{1}\), \(P_{2}\)). 
The fiducial, massless neutrino model is shown in solid blue, the 
\qty{0.1}{\electronvolt} neutrino model in dashed green, and the 
\qty{0.4}{\electronvolt} model in dotted orange. Notably,  \(P_{1}\)
pairs (saddle-1 to saddle-2) are the most sensitive topological probes of the sum of neutrino masses, especially at high redshift.}
\label{fig:particles-apex}
\end{figure*}

To complement the analysis of the persistence diagrams derived from \Quijote{} \gls{dm}  
density fields, we examine the apex of 
the most significant persistence pairs, which captures the characteristic densities of prominent topological features.
In this work, we define the apex as the mean position---i.e., the mean \(\log\) birth 
and death densities---of the top \qty{0.27}{\percent} of pairs in a persistence diagram ranked by the number of \(\sigma\)
assigned by \DisPerSE{}. This fraction corresponds to the probability beyond three standard deviations in a normal distribution, though the choice is arbitrary.
We compute the apex for each \Quijote{}  
snapshot, and in \cref{fig:particles-apex}, we show the mean apex positions and covariances over the \num{100} realizations per given cosmology. 
The left, middle, and right panels correspond to \(P_{0}\), \(P_{1}\), and \(P_{2}\)  pairs, respectively, while the top, middle, and bottom rows correspond to 
\(z = \num{2}\), \(\num{1}\), and \(\num{0}\). Each panel includes the three models: fiducial massless neutrinos (solid blue), \qty{0.1}{\electronvolt} massive neutrinos (dashed green), 
and \qty{0.4}{\electronvolt} massive neutrinos (dotted orange). Each point represents the mean apex across realizations, 
and the ellipse indicates the one-sigma covariance assuming a bivariate normal distribution.

The apex is clearly separated between the fiducial model and the heaviest neutrino case across all redshifts and pair types. 
For the lighter \qty{0.1}{\electronvolt} neutrino cosmology at \(z = 0\), the mean apex positions remain distinct, but their covariance ellipses overlap for 
\(P_{0}\) and \(P_{2}\) pairs, even encompassing each other's mean. This overlap decreases with increasing redshift. For these same pair types, 
the ellipses are elongated along the direction of the apex shift with increasing neutrino mass. 
In contrast, for \(P_{1}\) pairs, the overlap between the fiducial and \qty{0.1}{\electronvolt} models is already less pronounced at \(z = 0\), 
and the ellipses are elongated in a direction orthogonal to the apex shift with increasing 
neutrino masses. At \(z = 2\), the \(P_{1}\) ellipses are fully separated, with no overlap. 
Overall, \(P_{1}\) pairs (saddle-1 to saddle-2) appear to be the most sensitive topological probes of the sum of neutrino masses, particularly at high redshift, 
as also previously noted in \citet{Rossi:2022} and in \citet{Moon:2023}.

\subsection{Betti Curves}        \label{subsec:results-snapshots-betti-curves}

\begin{figure}[htp!]
\centering 
\includegraphics[trim=24 2 56 41, clip, width=\columnwidth]{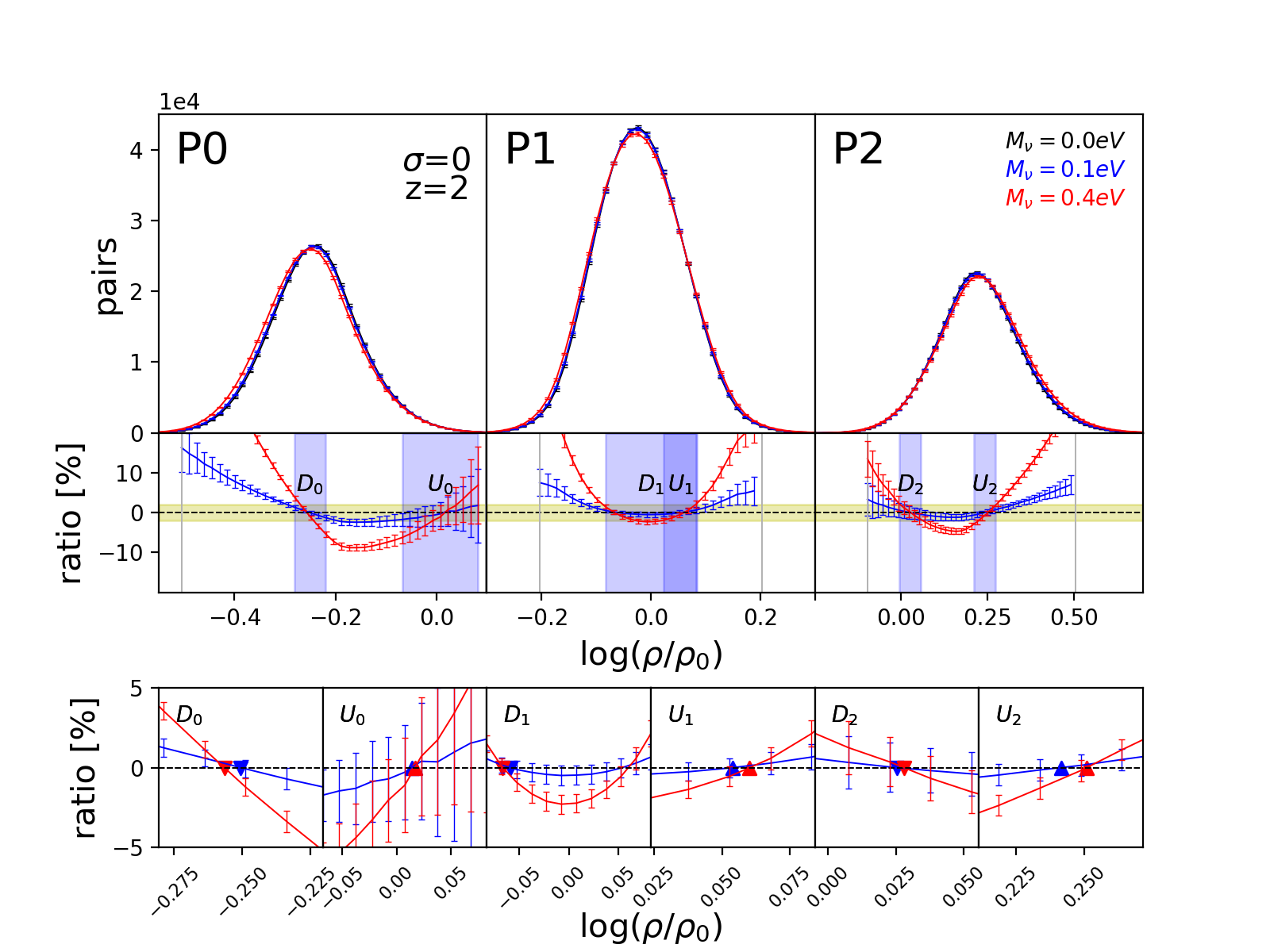}  
\includegraphics[trim=24 2 56 41, clip, width=\columnwidth]{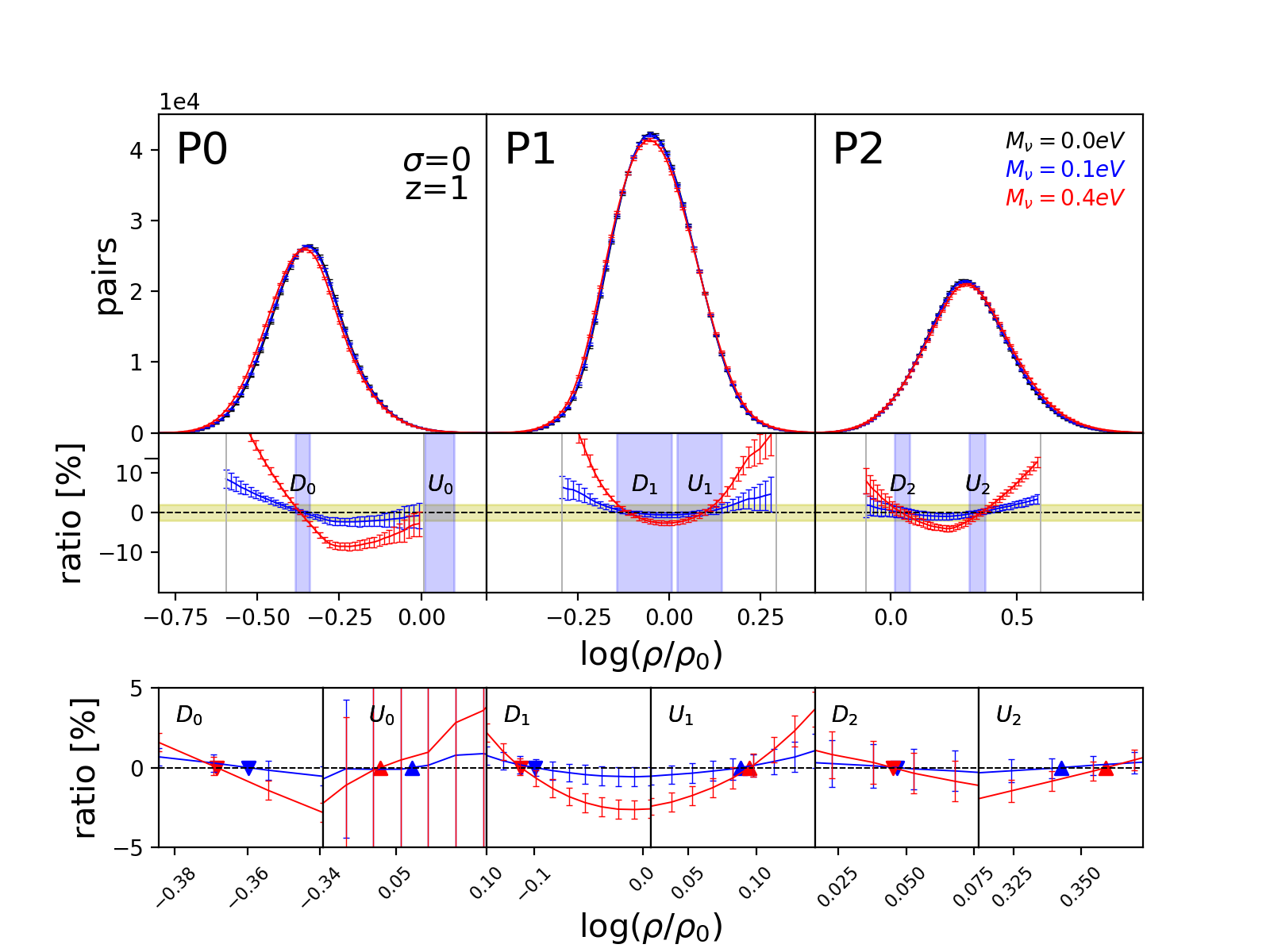}
\includegraphics[trim=24 2 56 41, clip, width=\columnwidth]{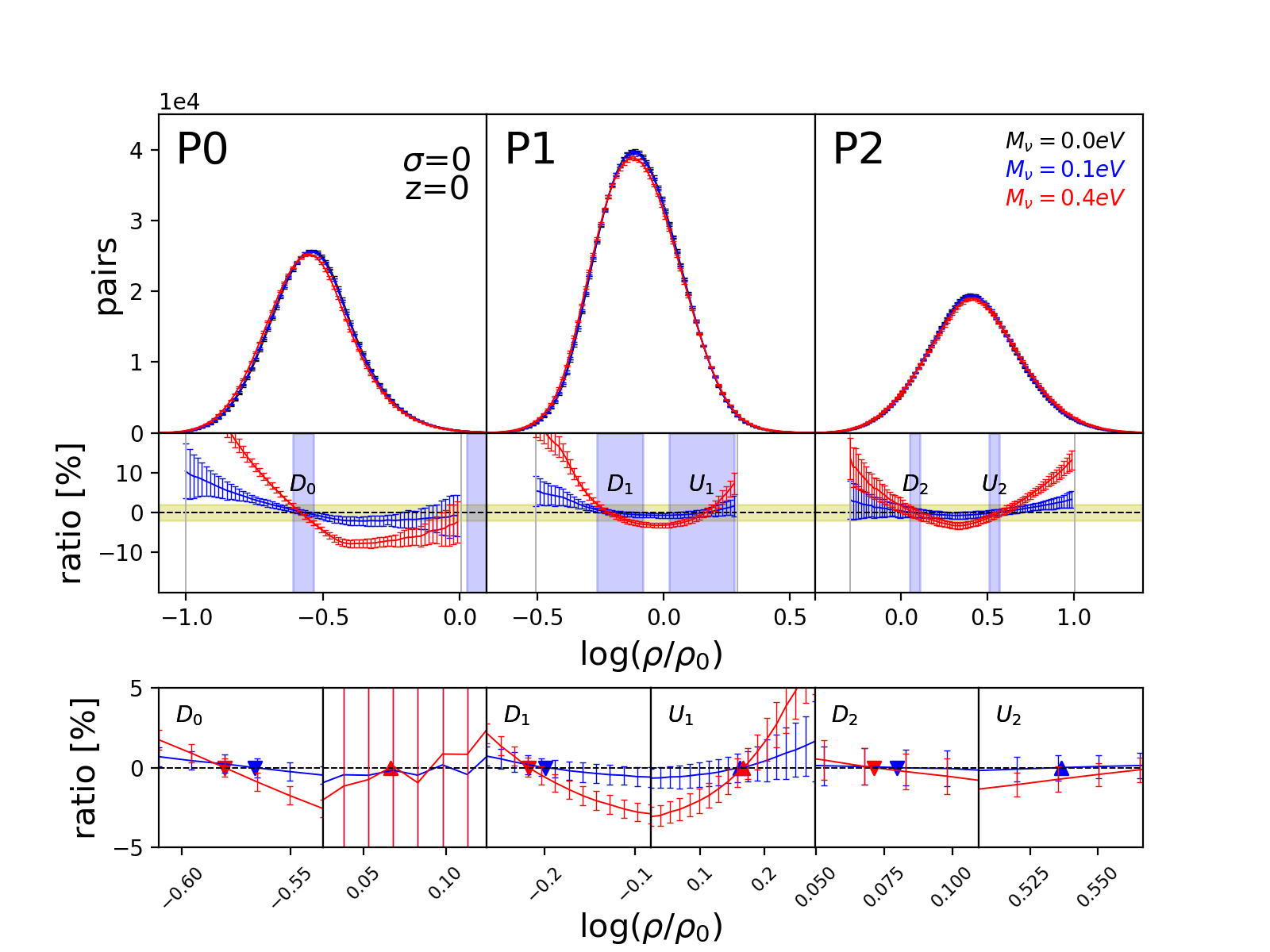}
\caption{Mean Betti curves computed from \Quijote{} \gls{dm}  
density fields for each simulation set at 
\(z = \numlist{2;1;0}\) (top to bottom) and for \(P_{0}\), \(P_{1}\) and \(P_{2}\) (left to right). 
Black lines show the fiducial massless neutrino case, blue lines the \qty{0.1}{\electronvolt} scenario, 
and red lines the \qty{0.4}{\electronvolt} model. In each block, the middle-row inset shows the ratio relative to the fiducial case, 
and the bottom-most panel provides a zoom-in on the Betti curve crossing points highlighted in the ratio panels with a blue shade. 
No persistence pair simplification is applied  (\(n\sigma = 0\)). See the main text for details.}
\label{fig:particles-betti}
\end{figure}

\begin{figure}[htp!]
\centering
\includegraphics[trim=14 19 56 42, clip, width=\columnwidth]{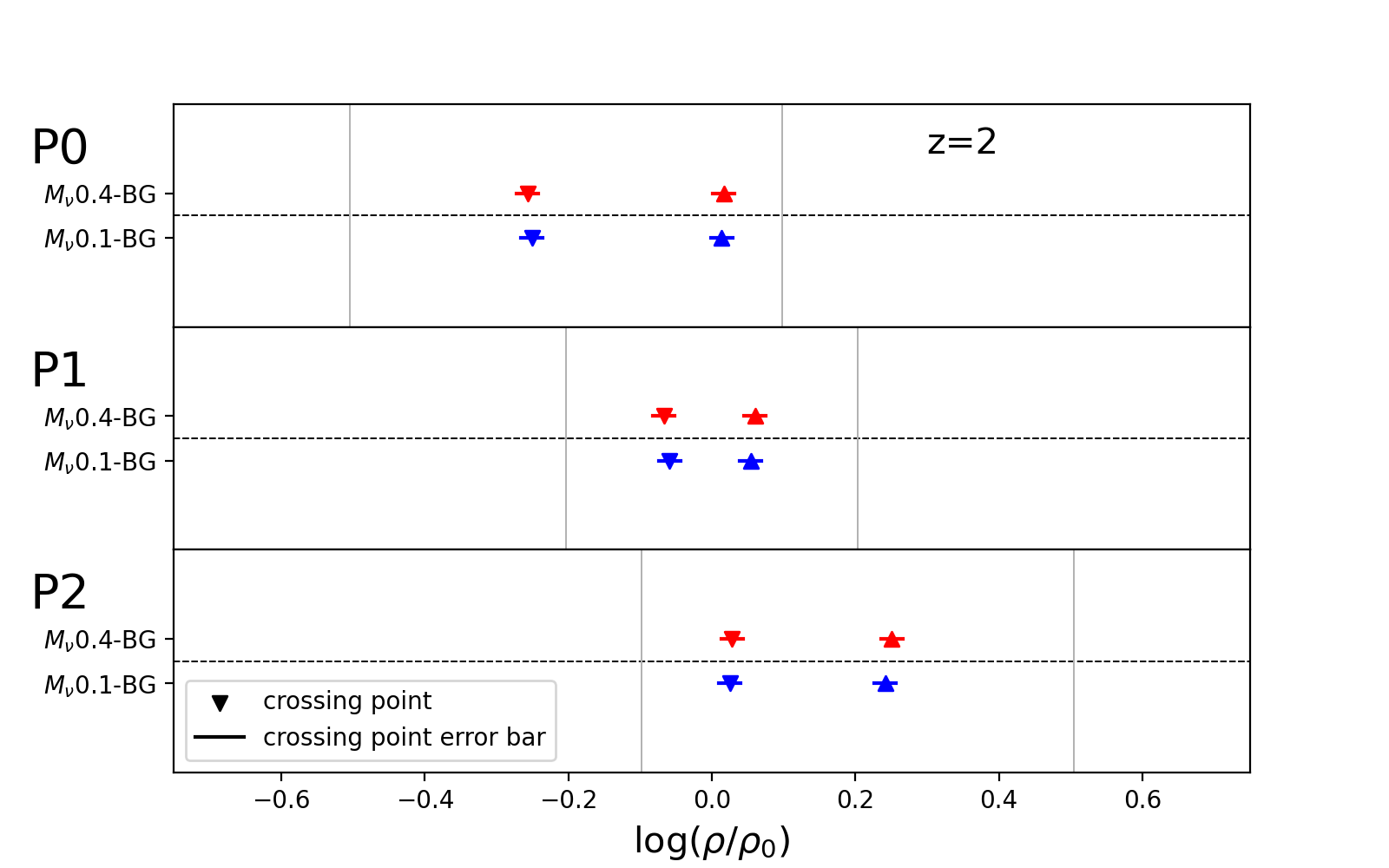}
\includegraphics[trim=14 19 56 42, clip, width=\columnwidth]{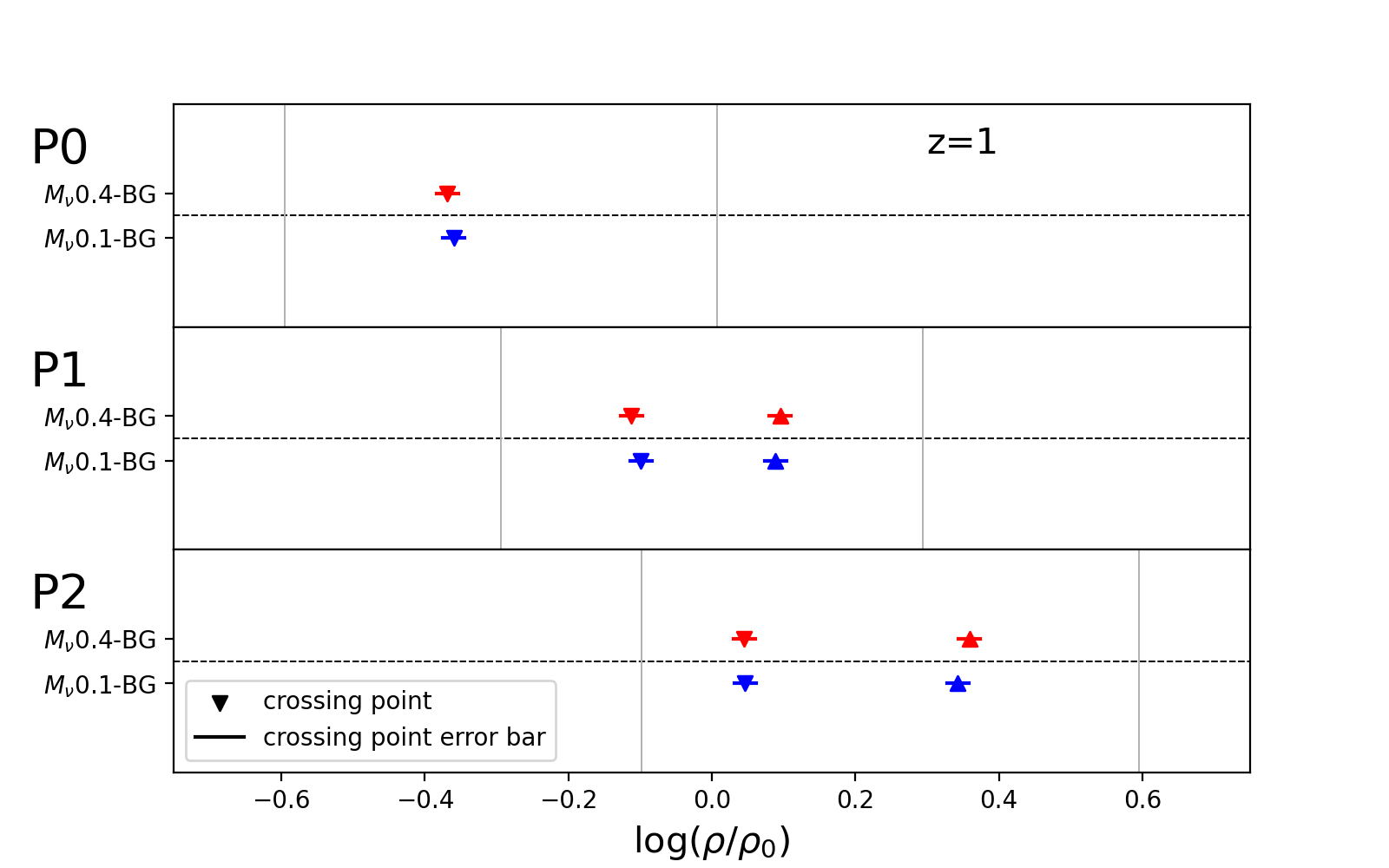}
\includegraphics[trim=14  2 56 42, clip, width=\columnwidth]{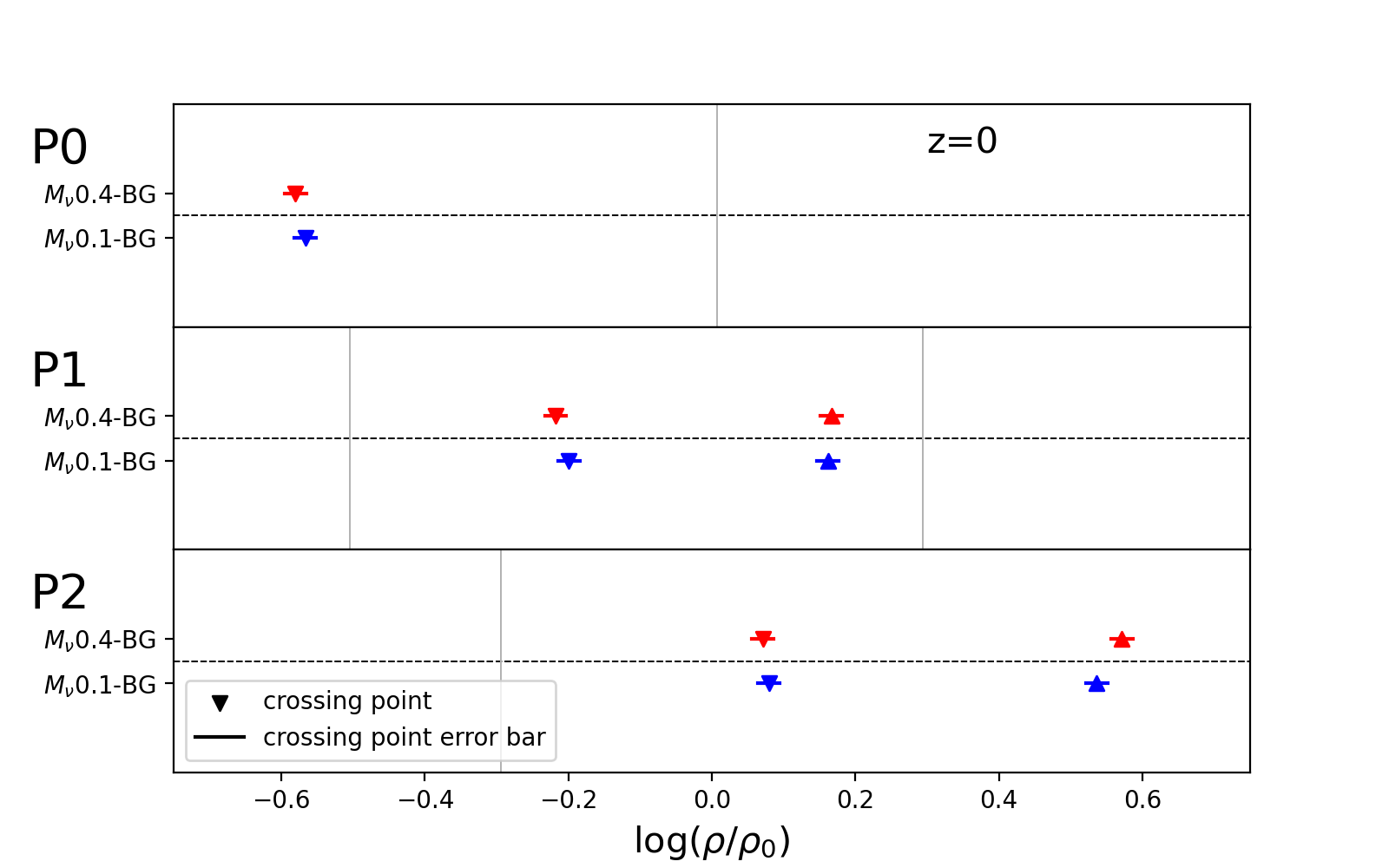}
\caption{Positions of  the Betti curve relative-difference crossing points derived from the \Quijote{} \gls{dm}  
fields, as shown in \cref{fig:particles-betti}, at 
\(z = \numlist{2;1;0}\) (top to bottom).  Downward and upward crossings are displayed as down- and up-pointing triangles, with the 
\(1\sigma\)  error on the mean indicated for each point. 
The top, middle, and bottom rows in each panel correspond to the \(P_{0}\), \(P_{1}\), and \(P_{2}\) curves, respectively. 
Crossing points for massive neutrino simulations with \qty{0.1}{\electronvolt}
are displayed in blue, and those with \qty{0.4}{\electronvolt} in red.}
\label{fig:particles-betti-crossing}
\end{figure} 

Building on the persistence diagrams presented above, we compute the Betti curves from \gls{dm}   
density fields to quantify the number of topological features as a function 
of the density threshold, providing a complementary view of the cosmic web structure.

In \cref{fig:particles-betti}, we show the Betti curves computed from \Quijote{} \gls{dm}  
density fields without simplification 
(i.e., counting all persistence pairs regardless of their associated number of  \(\sigma\)) for
\(z=2\) in the top block,  \(z=1\) in the middle, and  \(z=0\)
at the bottom. In each block, the top row displays the mean Betti curves for 
\(P_{0}\),  \(P_{1}\), and  \(P_{2}\) in their respective columns, while the middle row shows the relative ratios of the 
Betti curves with respect to the fiducial, massless neutrino simulations. 
Error bars indicate the standard deviation of the mean over \num{100} realizations per cosmology, and the
\(\pm\qty{2}{\percent}\)  region  is highlighted in yellow.
The bottom row of each block presents zoom-ins of the relative ratios (see below for details).

We find that massive neutrinos reduce the Betti numbers in the density range around the peaks of the Betti curves. 
The suppression is strongest for \(P_{0}\) pairs, with a decrease of  \(\approx\qty{2}{\percent}\) 
for \qty{0.1}{\electronvolt} neutrinos and  \(\approx\qty{10}{\percent}\) for \qty{0.4}{\electronvolt}. 
The effect is smallest for  \(P_{1}\) pairs, remaining below  \(\qty{1}{\percent}\)
for \qty{0.1}{\electronvolt} and reaching \(\approx\qty{2}{\percent}\) for \qty{0.4}{\electronvolt}. 
Regarding \(P_{2}\) pairs, the reduction is intermediate, at \(\approx\qty{1}{\percent}\)
for \qty{0.1}{\electronvolt} and \(\approx\qty{5}{\percent}\) for \qty{0.4}{\electronvolt}. 
 
Another key feature of \cref{fig:particles-betti} is the crossing of the Betti curves. 
Here, a crossing is defined from the ratio of each massive neutrino Betti curve to the baseline massless neutrino cosmology, 
where the horizontal line corresponds to the fiducial model (middle and bottom rows in each block of the figure). 
Each ratio exhibits two crossing points: a ``downward'' crossing, where the massive neutrino Betti curve falls below the fiducial, 
and an ``upward'' crossing, where it rises above. 
The suppression of Betti numbers in massive neutrino cosmologies occurs in a density range around the curve peak, 
while outside this range the effect is reversed, showing a neutrino mass-dependent increase. 
For each pair type, the two Betti curve crossing points are generally similar, regardless of the neutrino mass. 
Zoom-ins of these crossing points are shown in the third row of each block in \cref{fig:particles-betti}, with the 
vertically highlighted regions in the ratio panels labeled `{D}' for ``downward'' and `{U}' for ``upward'' corresponding to the zoom-in panels below. 
For each pair of crossing points---one for \qty{0.1}{\electronvolt} and the other for \qty{0.4}{\electronvolt}---the differences are subtle and less pronounced 
(though still observable) when considering the error bars. 

We verify that the Betti curve crossing points are largely insensitive to neutrino mass in \cref{fig:particles-betti-crossing}, 
which presents the mean positions of the ``downward'' (lower triangles) and ``upward'' (upper triangles) Betti curve crossings, 
along with the standard deviation of the mean over  \num{100} simulations per cosmology, across the three redshift intervals considered. 
The crossings correspond to intersections between the fiducial, massless neutrino simulations 
and those with \qty{0.1}{\electronvolt} (blue) and \qty{0.4}{\electronvolt} (red) neutrinos, respectively.
For  \(P_{0}\), \(P_{1}\), and \(P_{2}\), 
the respective ``downward'' and ``upward'' crossings shift to higher densities, reflecting the ordering of the Betti curve peaks, with 
\(P_{0}\)  on the lower density side and  \(P_{2}\) 
on the higher density side. 
 
Interestingly, most crossing points in \cref{fig:particles-betti-crossing} appear at similar positions regardless of neutrino mass. The upward crossing of 
\(P_{2}\) pairs (saddle-2 to peaks) shifts toward higher densities more rapidly for larger neutrino masses, with the two points clearly separated at 
\(z=0\). For \(P_{0}\) pairs (void to saddle-1), a clear upward crossing is absent, as it occurs only near the edge of the Betti curves where 
all curves converge to zero and the error bars become large.  

All the mass-dependent signatures highlighted in this section are detectable at the few-percent level, even for 
$M_\nu \sim 0.1$ eV, providing a robust and physically grounded probe of massive neutrinos in the cosmic web.

 
\section{Results from Dark Matter Halos}      \label{sec:results-halos}

We now turn to the analysis of dark matter halos, presenting the topological properties derived from the halo catalogs, 
including persistence diagrams and Betti curves, and comparing them with the baseline, massless neutrino cosmology. 
Because the number of points in halo catalogs is relatively small, we can directly use the methodology 
detailed in \cref{subsubsec:density-from-halos}, 
employing \gls{dtfe} to triangulate the distributions. In what follows, we show the results for the same neutrino mass 
models considered previously (\qty{0.1}{\electronvolt} and \qty{0.4}{\electronvolt}).

\subsection{Persistence Diagrams}        \label{subsec:results-halos-persistence-diagrams}

\begin{figure}[htp!]
\centering                                                                                                                                                                                            
\includegraphics[trim=31 51 19 73, clip, width=0.96\columnwidth]{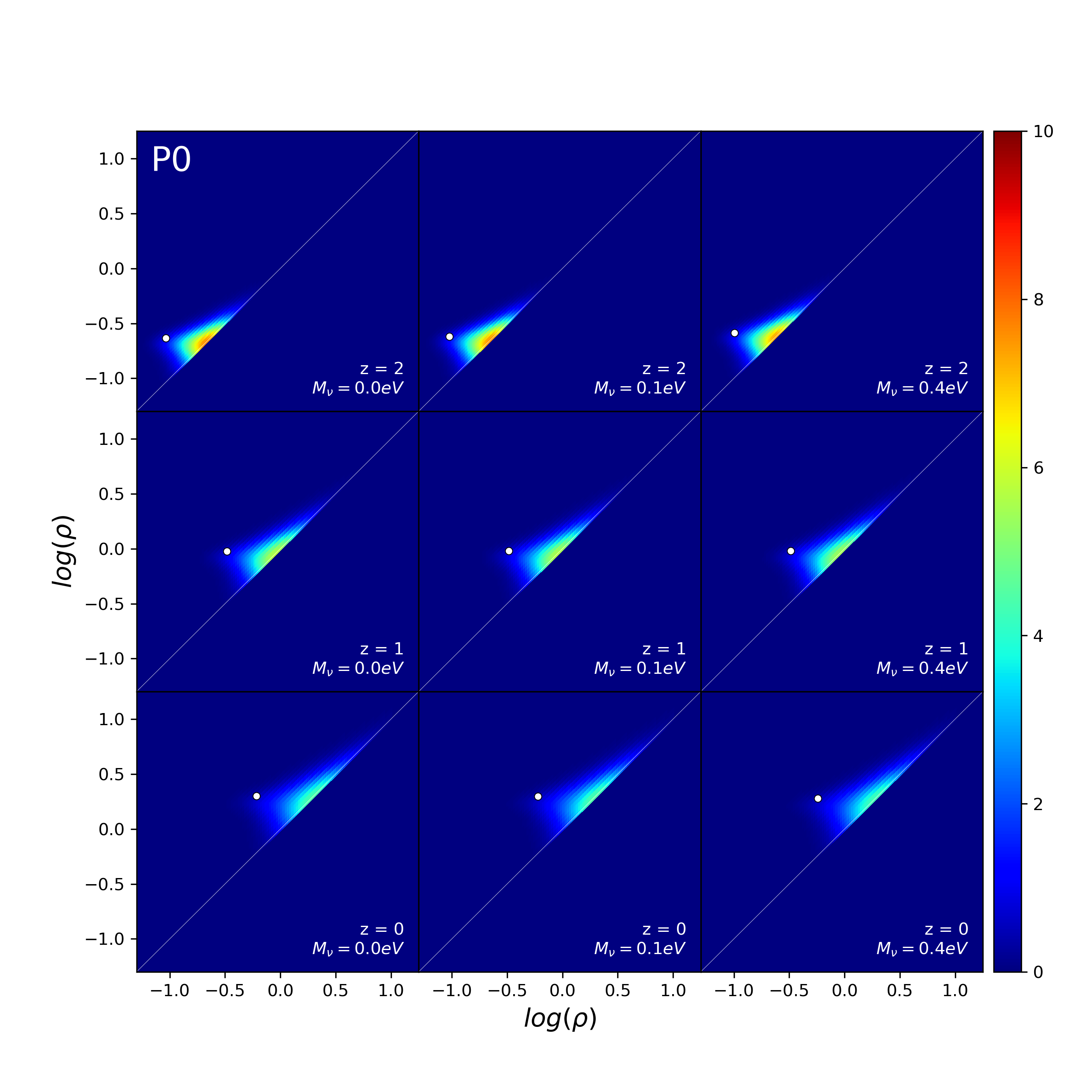}
\includegraphics[trim=31 51 19 73, clip, width=0.96\columnwidth]{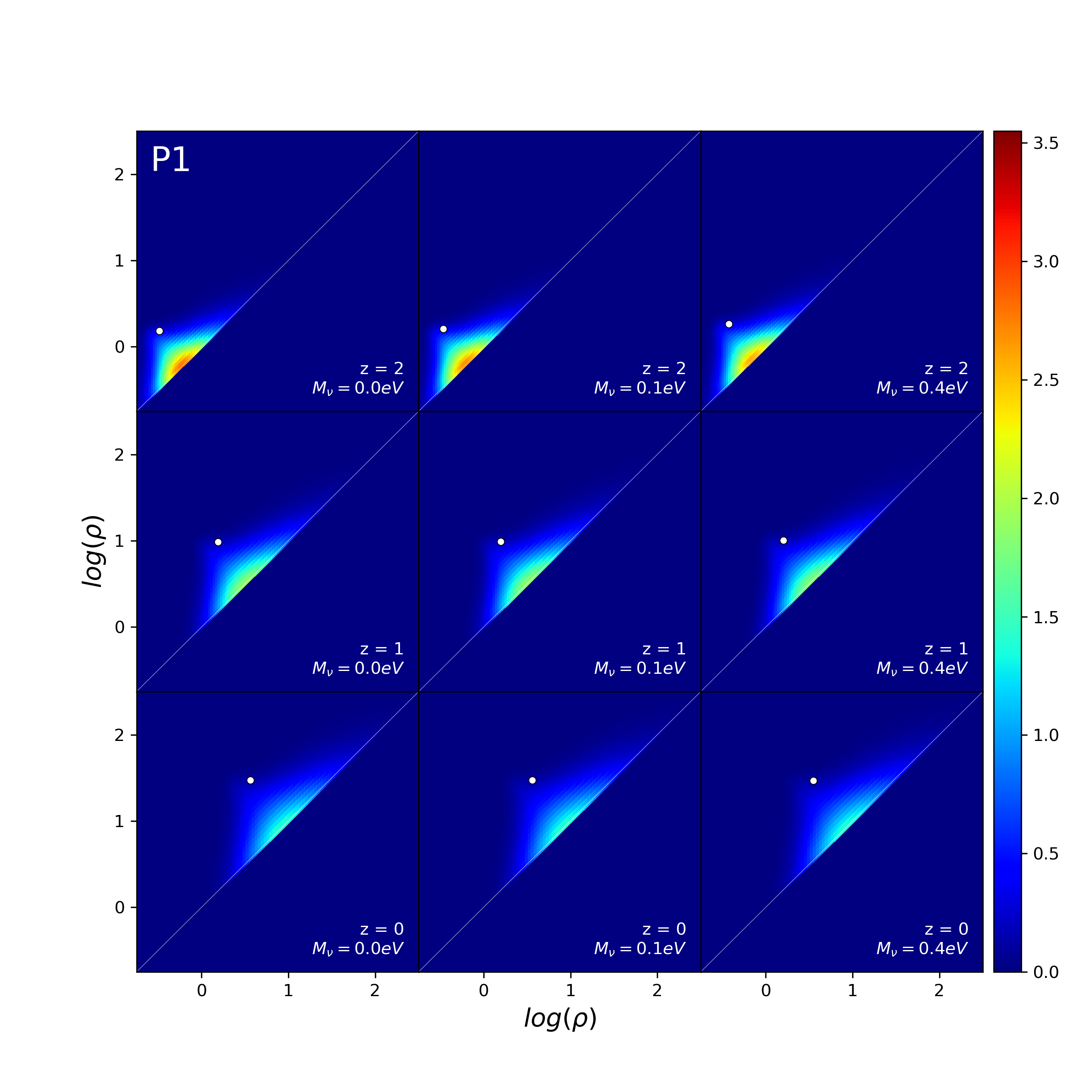}
\includegraphics[trim=31 35 19 73, clip, width=0.96\columnwidth]{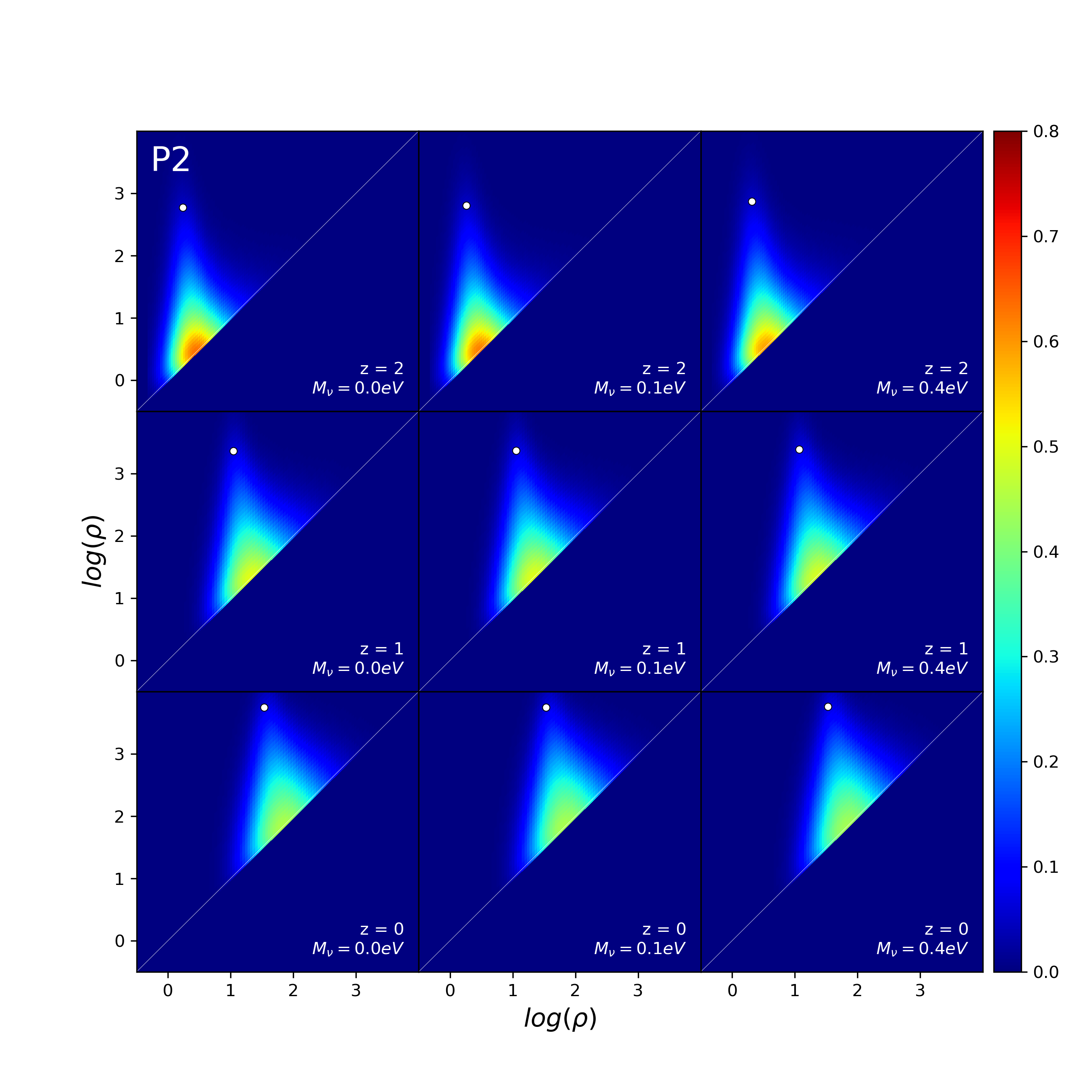}
\caption{Persistence diagrams for the \Quijote{} halo mass density. The layout differs slightly from \cref{fig:particles-persistence}. 
Blocks from top to bottom correspond to the pair types \(P_{0}\), \(P_{1}\) and \(P_{2}\).
Within each block, rows from top to bottom represent redshifts \(z = \numlist{2;1;0}\),
and columns from left to right correspond to the \Quijote{} 
fiducial (massless neutrinos), \qty{0.1}{\electronvolt}, and \qty{0.4}{\electronvolt} neutrino cosmologies. 
White dots indicate the apex positions (see \cref{fig:halo-apex}).}
\label{fig:halos-persistence}
\end{figure}

We present the persistence diagrams for \Quijote{} halos in \cref{fig:halos-persistence}, which are 
analogous to those in \cref{fig:particles-persistence} 
but arranged in a slightly different layout. In the figure, the top, middle, and bottom blocks correspond to 
\(P_{0}\), \(P_{1}\), and \(P_{2}\)  pairs, respectively. 
Within each block, the rows (from top to bottom) represent redshifts 
\numlist[list-final-separator={, and }]{2;1;0}, while the columns 
(from left to right) correspond to the \Quijote{} fiducial, massless neutrino cosmology, 
the \qty{0.1}{\electronvolt} massive neutrino model, and the \qty{0.4}{\electronvolt} massive neutrino scenario.

The overall triangular shapes resemble those observed for \gls{dm}   
particles in \cref{fig:particles-persistence},
with \(P_{0}\)  pairs skewed toward lower densities, \(P_{2}\) toward higher densities, and and \(P_{1}\) 
more centered. However, the edges are generally less well-defined than in the \gls{dm}  
persistence diagrams. 
In particular, the  \(P_{1}\) diagrams for halos largely lose their triangular shape at \(z = 0\), 
adopting instead a disk-segment-like form. Signs of phase transitions in the 
sublevel set remain visible---especially at high redshift---but the transitions appear more gradual.

\begin{figure}
\centering
\includegraphics[trim=19 88 54 60, clip, width=\columnwidth]{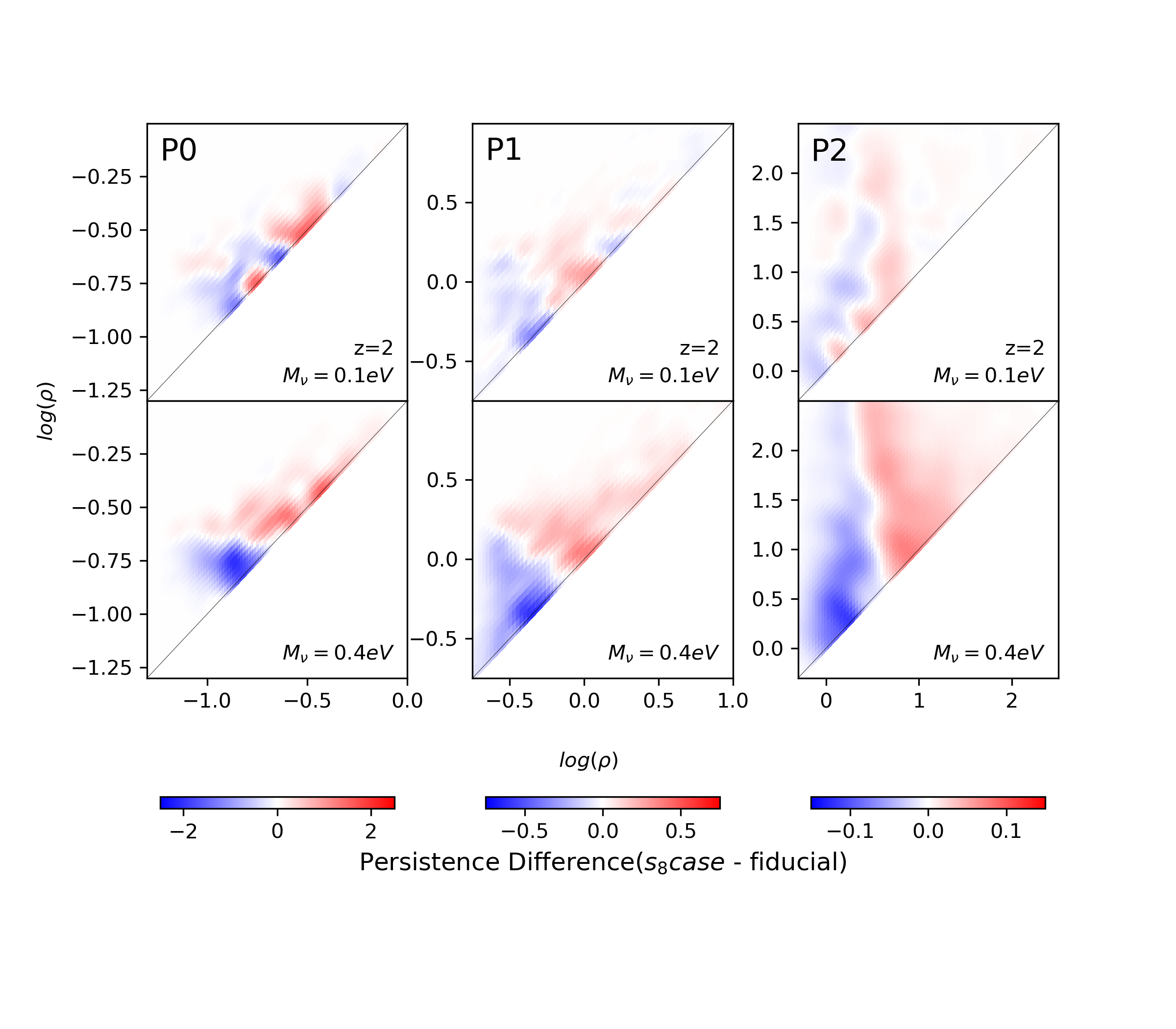}
\includegraphics[trim=19 88 54 60, clip, width=\columnwidth]{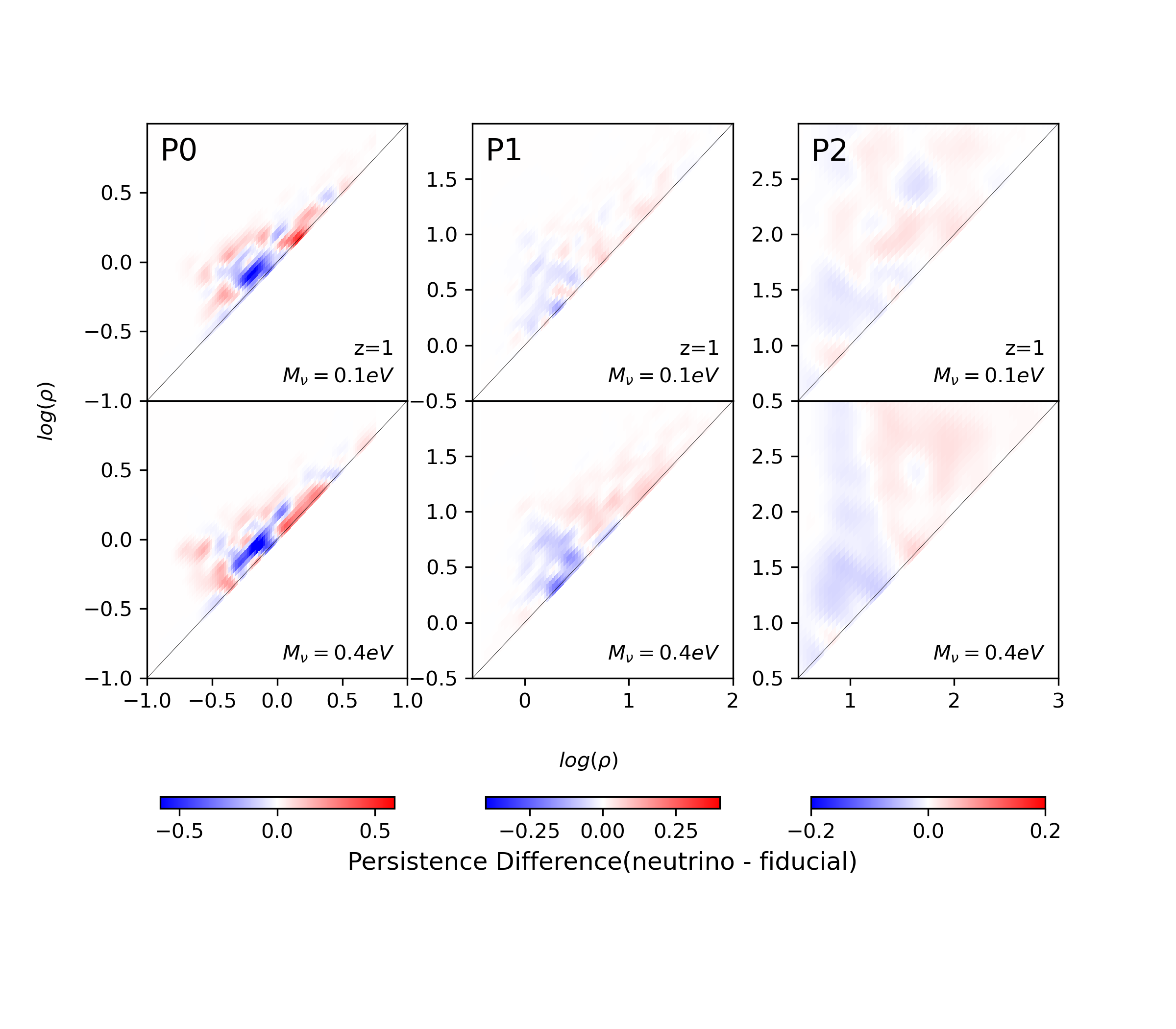}
\includegraphics[trim=19 77 54 60, clip, width=\columnwidth]{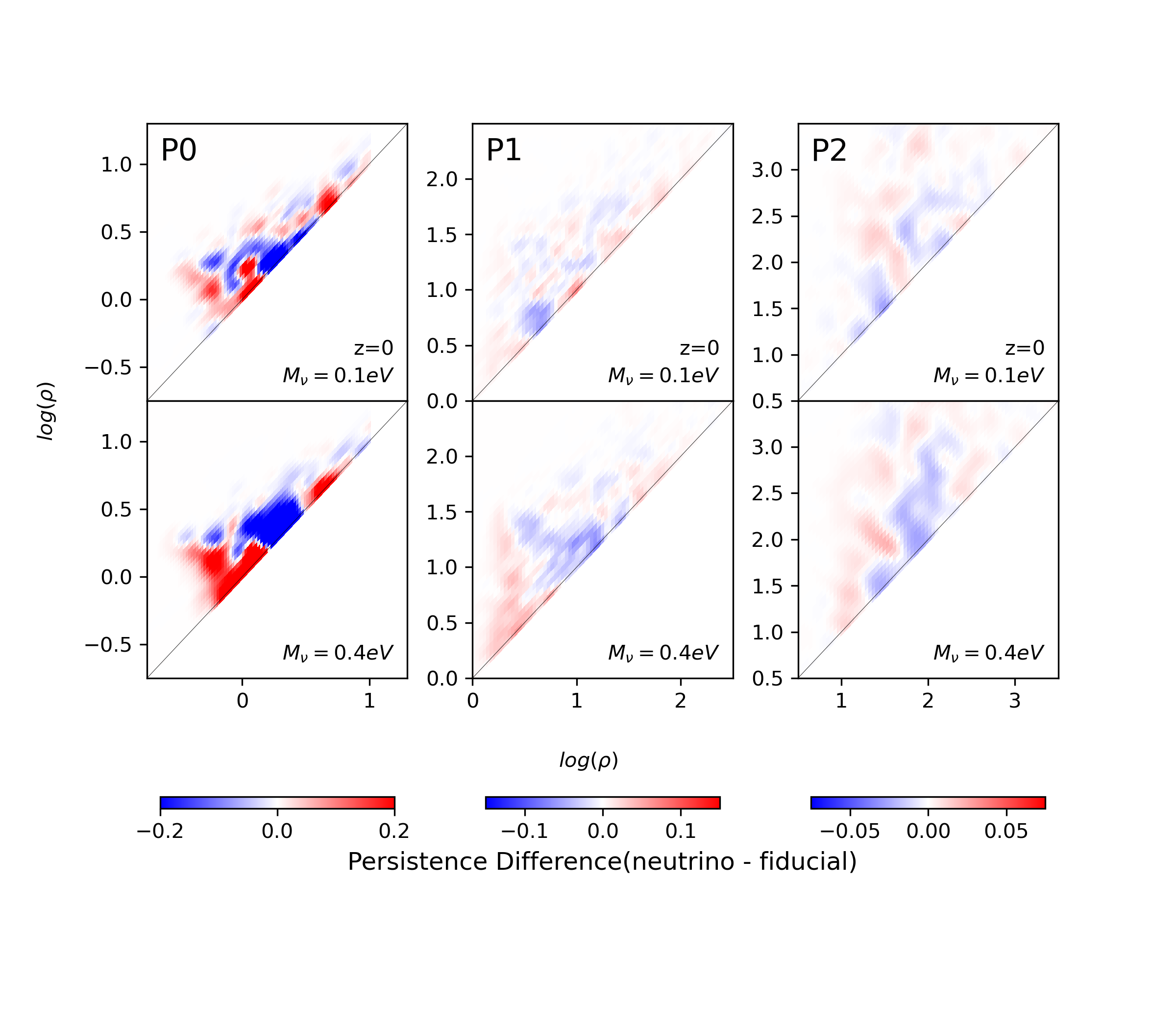}
\caption{Persistence diagram differences for the \Quijote{} halo density fields at 
\(z=2\) (top block), \(z=1\) (middle block), and \(z=0\) (bottom block),
showing the change in pair number density between the massive neutrino simulations 
and the fiducial massless neutrino case. Within each block, columns correspond to pair types: 
\(P_{0}\) (left), \(P_{1}\) (middle), and \(P_{2}\) (right). 
The layout follows \cref{fig:particles-persistence-difference}. See the main text for details.}
\label{fig:halo-persistence-difference}
\end{figure}

In \cref{fig:halo-persistence-difference}, we show the differences in pair number density in the \Quijote{} halo persistence diagrams, 
analogous to \cref{fig:particles-persistence-difference} for the \gls{dm}  
case.  Each block corresponds to a redshift: top  \(z=2\),
middle  \(z=1\), and bottom  \(z=0\).
Columns indicate pair types: left \(P_{0}\), middle \(P_{1}\),  
and right \(P_{2}\). Within each block, the top row displays the 
difference for \qty{0.1}{\electronvolt} neutrinos relative to the fiducial massless neutrino model, 
and the bottom row shows the \qty{0.4}{\electronvolt} case relative to the fiducial.

\begin{figure*}[htp!]
\centering
\includegraphics[angle=0,width=0.93\textwidth]{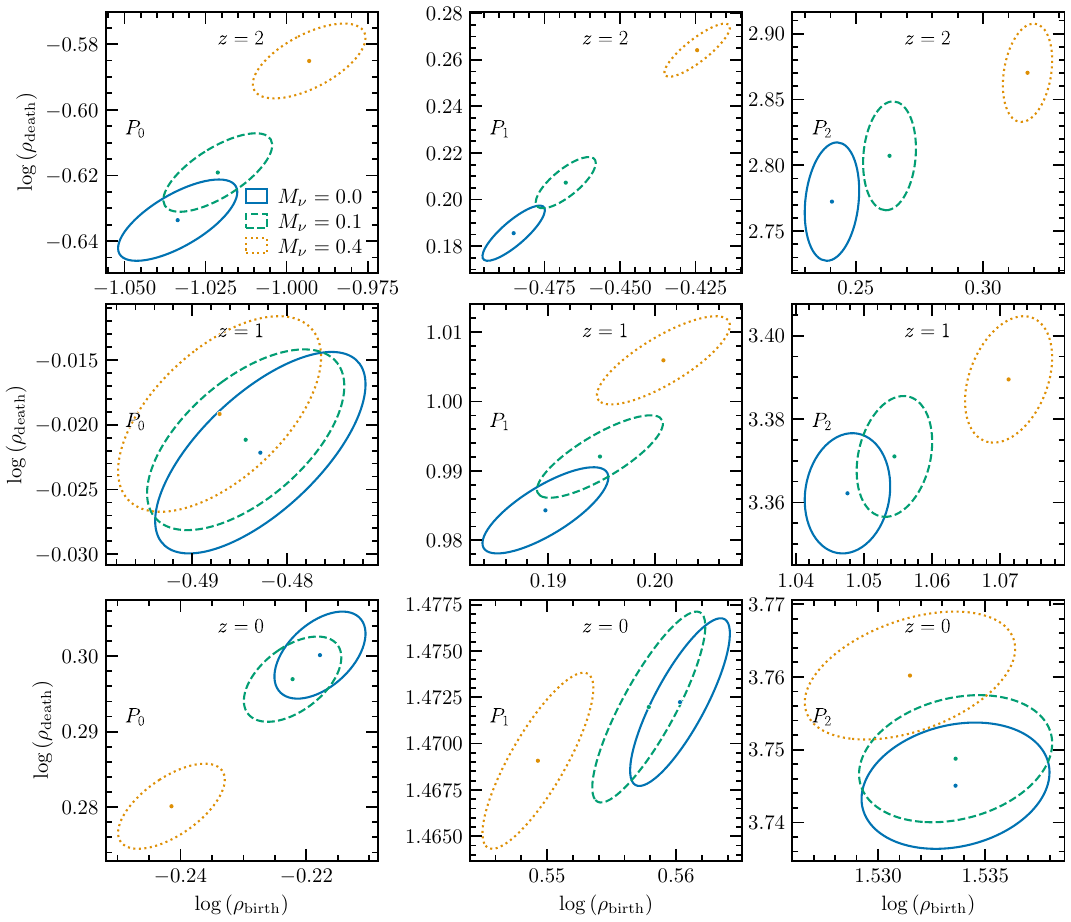}
\caption{Mean positions (points) of the persistence diagram apexes for the \Quijote{} halo mass density fields, with ellipses indicating the 
\num{1}\(\sigma\) variation over \num{100} simulations per cosmology. The layout mirrors \cref{fig:particles-apex}, with rows corresponding to redshifts 
\(z = \numlist{2;1;0}\) and columns to pair types \(P_{0}\), \(P_{1}\) and \(P_{2}\).  Differences in apex positions are more pronounced at higher redshift,
particularly for  \(P_{1}\) (saddle-1 to saddle-2) and  \(P_{2}\) (saddle-2 to maximum) pairs.} 
\label{fig:halo-apex}
\end{figure*}

At  \(z=2\), persistence pairs for massive neutrino simulations shift toward higher densities for all pair types, 
most clearly for \qty{0.4}{\electronvolt}, while the \qty{0.1}{\electronvolt} case shows the same 
trend but with a noisier difference diagram, 
indicating a stronger shift for larger neutrino masses. 
In contrast to the \gls{dm}   
case, all pair types at  \(z=2\) 
shift in the same direction. As redshift decreases, the difference diagrams become noisier, 
showing that these shifts weaken over time. Some differences also reverse 
with redshift;  
for example,  \(P_{0}\) pairs shift  with increasing neutrino masses to higher densities at \(z=2\)
but toward lower densities at  \(z=0\), 
with no clear trend at  \(z=1\)
due to noise. These patterns are consistent with 
the mean apex movements shown in \cref{fig:halo-apex}, as we discuss next.

As with the \gls{dm}  
persistence diagrams, we estimate the apex position for each halo simulation as the 
mean birth and death densities of the persistence pairs with the highest number of \(\sigma\).
For halos, we increase this fraction to \qty{4.6}{\percent} (up from \qty{0.27}{\percent} in the \gls{dm}  
case), corresponding to the probability 
outside two standard deviations of a normal distribution (compared with three standard deviations for \qty{0.27}{\percent}). 
This choice is somewhat arbitrary, but because halo persistence diagrams contain far fewer pairs, we find that a larger selection 
fraction yields a more robust apex estimate. The mean apex computed over 100 realizations per cosmology 
is shown as a white point on each persistence diagram in \cref{fig:halos-persistence}.

The mean apex positions obtained from the \Quijote{} halo density fields, along with their covariance across \num{100} realizations per cosmology, 
are shown in \cref{fig:halo-apex}, providing the halo analog of \cref{fig:particles-apex} for the \gls{dm}  
field. Each row corresponds, from top to bottom, to redshifts \numlist[list-final-separator={, and }]{2;1;0}, 
and each column, from left to right, to \(P_{0}\), \(P_{1}\), and \(P_{2}\) pairs.
Points and ellipses indicate the mean and covariance of the apex positions across realizations. 
The fiducial massless neutrino case is shown in solid blue, the \qty{0.1}{\electronvolt} scenario in 
dashed green, and the \qty{0.4}{\electronvolt} cosmology in dotted orange.

Compared to the \gls{dm}  
case (\cref{fig:particles-apex}), we again observe ellipse overlaps at \(z = 0\), 
with the fiducial and \qty{0.1}{\electronvolt} mean apexes lying within each other's ellipses for all three pair types. 
Differences in apex positions are more pronounced at higher redshift, particularly for  \(P_{1}\) (saddle-1 to saddle-2) and 
\(P_{2}\) (saddle-2 to maximum) pairs. \Cref{fig:halo-apex} also highlights a transition between 
\(z = 2\)  and \(z = 0\), where the direction of apex movement with increasing neutrino mass reverses. 
For example, in the  \(P_{0}\) case, the order of the mean points is reversed between  \(z = 2\)
and  \(z = 0\), with all ellipses overlapping at  \(z = 1\).
This transition is consistent with the patterns seen in the difference diagrams of \cref{fig:halo-persistence-difference}.

Similar to the \gls{dm}   
persistence diagram apexes, the largest differences appear at the highest redshift. 
In the halo case, the ellipses for \(P_{2}\) pairs are all non-overlapping, with their major axes oriented roughly \ang{45} relative 
to the direction of apex movement with increasing neutrino mass. 
The \(P_{1}\) pair ellipses are also clearly distinct, but in this case their movement is aligned with the direction of their major axes. 

\subsection{Betti Curves}       \label{subsec:results-halos-betti-curves}

\begin{figure}[htp!]
\centering
\includegraphics[trim=6 8 5 7, clip, width=\columnwidth]{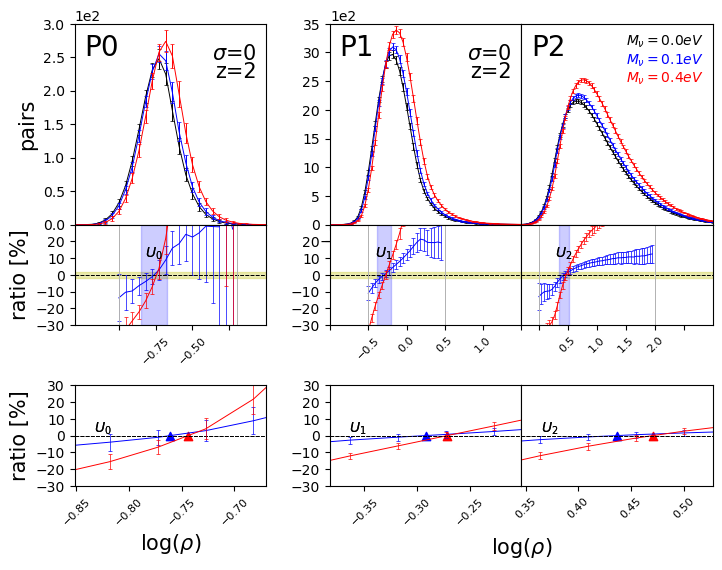}
\includegraphics[trim=6 8 5 7, clip, width=\columnwidth]{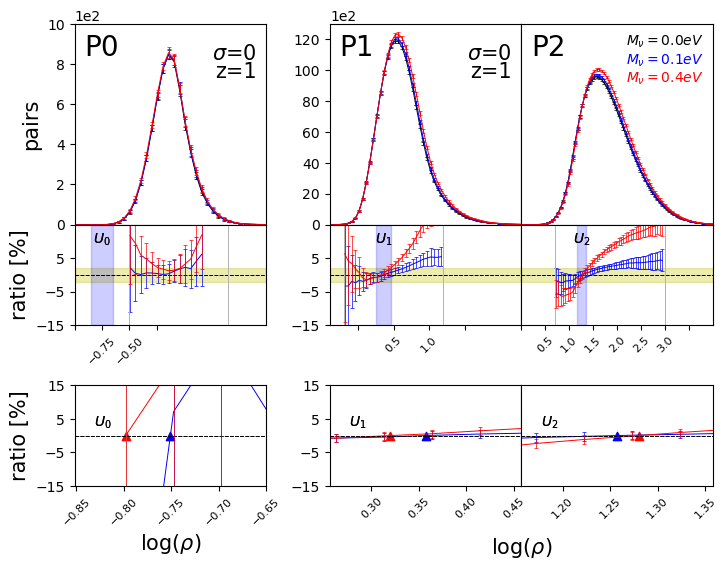}
\includegraphics[trim=6 8 5 7, clip, width=\columnwidth]{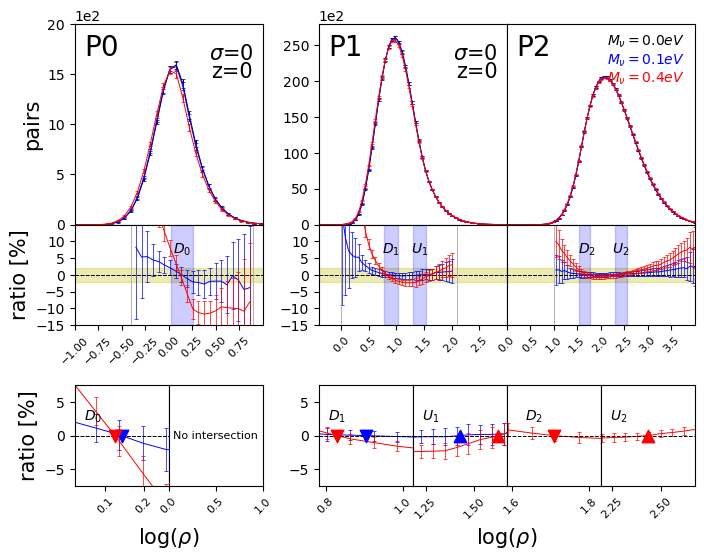}
\caption{Mean Betti curves for the \Quijote{} halo mass density fields. 
The figure is organized in three blocks, from top to bottom corresponding to \(z = \numlist{2;1;0}\), respectively.
Within each block, the top row shows the mean Betti curves for the three different cosmologies considered, 
the middle row presents their ratios relative to the fiducial, massless neutrino case, and the bottom row zooms in on the Betti curve crossing points. 
The layout mirrors that of \cref{fig:particles-betti}, allowing direct comparison with the \gls{dm}  
density field results.  Error bars indicate the standard deviation of the mean over \num{100} realizations per cosmology.}
\label{fig:halos-betti}
\end{figure}

We next present the Betti curves derived from the \Quijote{}  halo catalogs, using the same approach as for the \gls{dm}  
field. Despite the lower number of tracers, the curves retain clear topological signatures and enable a robust comparison 
across cosmologies with different neutrino masses. 
In what follows, we examine their redshift evolution, highlighting both the key deviations from the fiducial 
massless neutrino case and the differences relative to the corresponding \gls{dm}  
field.

We begin with \cref{fig:halos-betti}, the halo counterpart of \cref{fig:particles-betti}. 
The figure is organized into three blocks corresponding to  \(z=2\) (top), 
\(z=1\) (middle), and  \(z=0\) (bottom). Within each block, the top row shows the mean Betti curves for the different cosmologies, 
the middle row presents the ratios relative to the fiducial massless neutrino case, and the bottom row provides zoom-in panels 
around the Betti curve crossing points. The mean curves and associated uncertainties are computed from the \num{100} realizations of each set.
 
At \(z=2\), for all pair types, the Betti curves for massive neutrinos cross the fiducial curve upward at a single point; 
that is, they transition from a negative to a positive difference as the density threshold increases. 
No clear downward crossing is visible; if present, it occurs in the high-density tail where all curves converge to zero and is therefore not informative. 
The differences at \(z=2\) are significantly larger than in the \gls{dm}   
density field case.  Around the peak, Betti numbers are higher by about \qty{20}{\percent} for the \qty{0.4}{\electronvolt} case and 
by about \qty{10}{\percent} for \qty{0.1}{\electronvolt}, relative to the fiducial, massless neutrino cosmology. 
Based on the error bars, \(P_{2}\) pairs (saddle-2 to maximum) show the most distinct separation, followed by 
\(P_{1}\) pairs (saddle-1 to saddle-2). 
The  \(P_{0}\)  pairs (minimum to saddle-1) follow a similar trend, but their larger, overlapping uncertainties reduce the statistical significance.
 
As redshift decreases, the transition already observed in the persistence diagrams 
(\cref{subsec:results-halos-persistence-diagrams}) becomes evident in the Betti curves as well. 
For \(P_{0}\) pairs, the upward crossing seen at  \(z=2\) 
shifts to a downward crossing at  \(z=0\).  At  \(z=1\),  
the  \(P_{0}\) mean Betti curves for massive neutrinos remain slightly above the fiducial model, but the error bars largely overlap, 
indicating that the difference is not statistically significant. 
For \(P_{1}\) and \(P_{2}\) pairs, the positive differences on the high-density side of the Betti curves decrease, 
while the low-density side shifts from negative to positive. 
By \(z=0\), the differences exhibit two crossing points, going from positive to negative and back to positive as density increases. 
Overall, the shape of the Betti curve differences at \(z=0\) resembles that of the \gls{dm} \%gls{cdm} 
case (\cref{fig:particles-persistence-difference}), 
although halos show no upward crossing for 
\(P_{0}\) pairs at high densities.

\begin{figure}[htp!]
\centering
\includegraphics[trim=10 24 56 39, clip, width=\columnwidth]{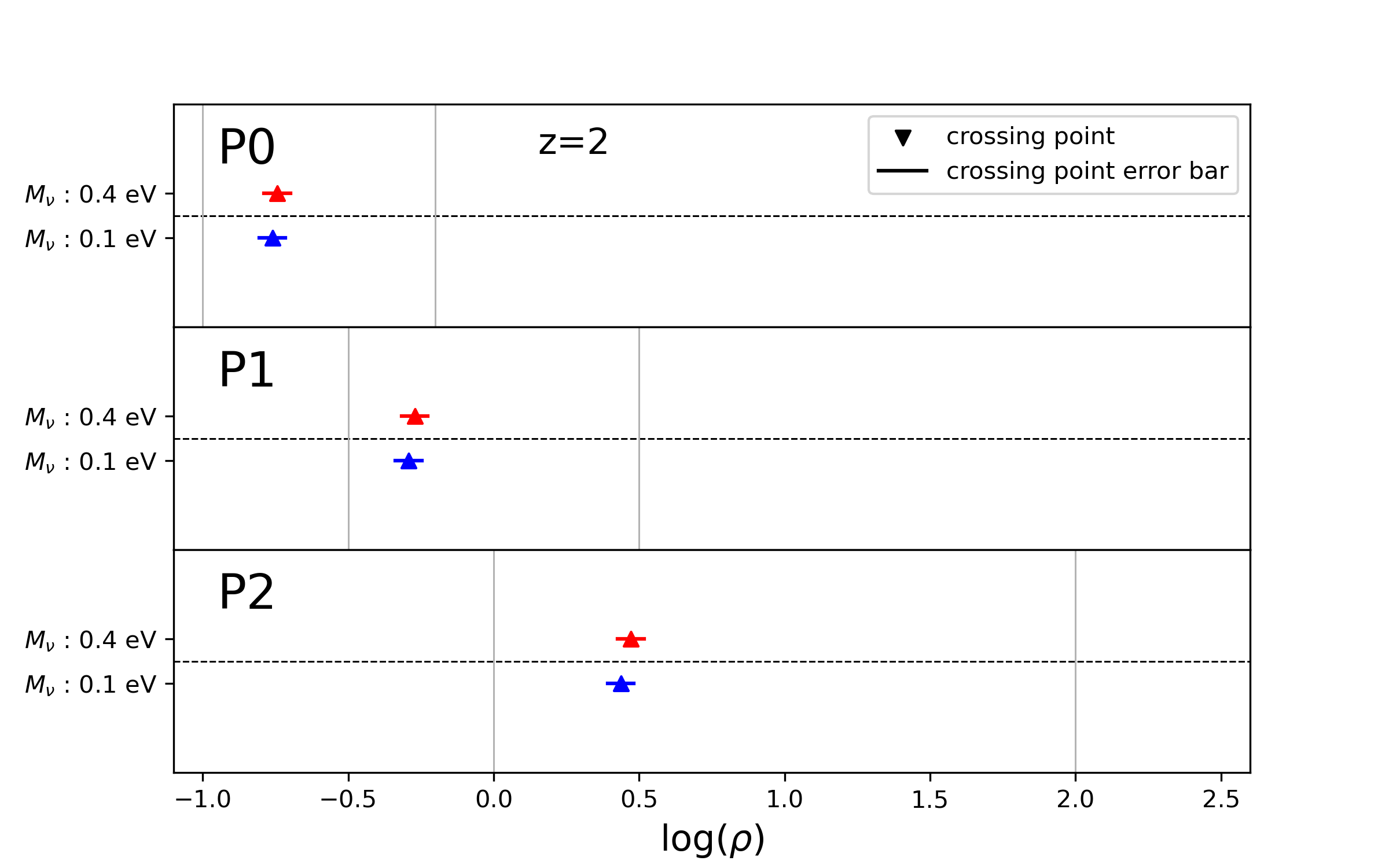}
\includegraphics[trim=10 24 56 39, clip, width=\columnwidth]{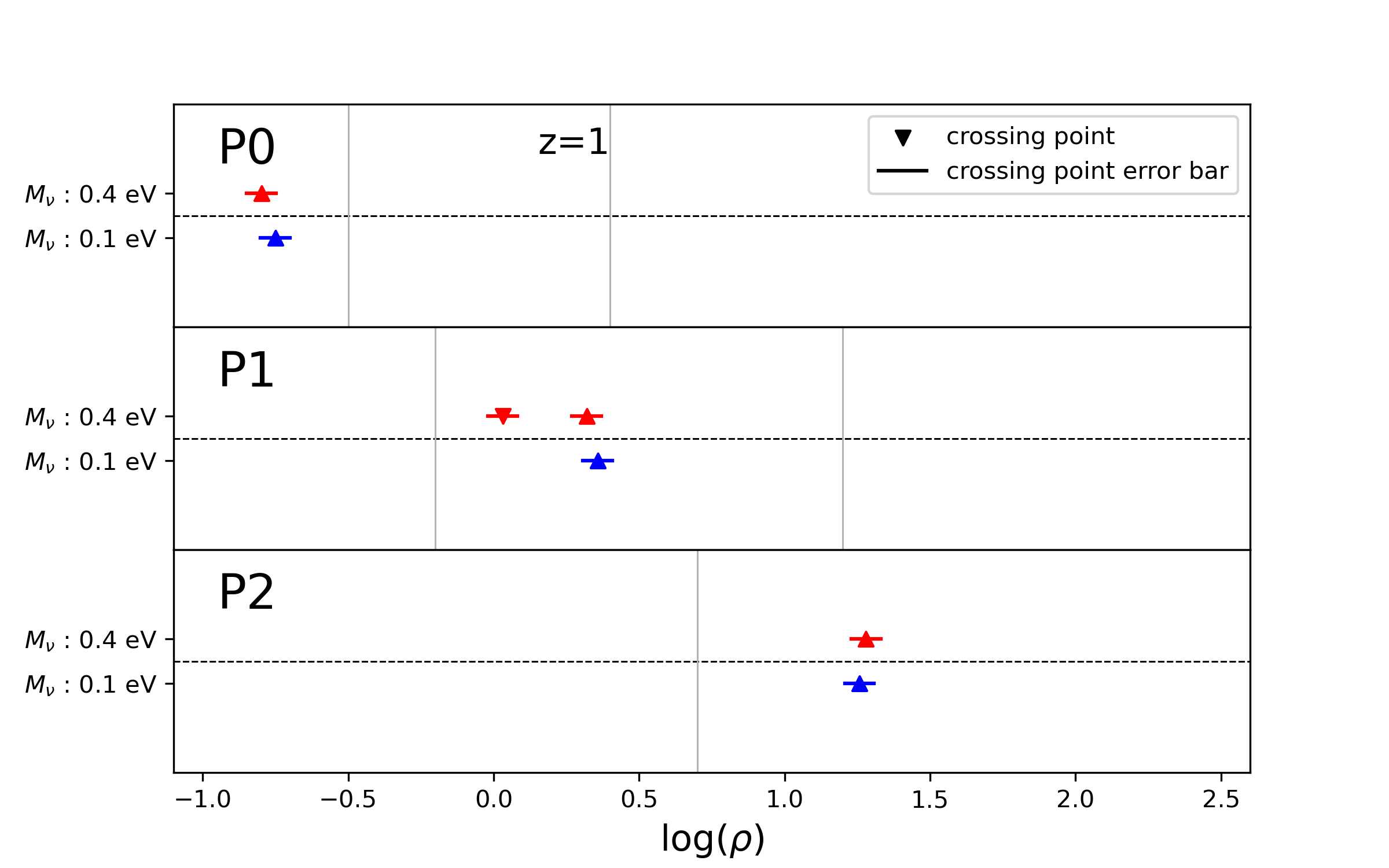}
\includegraphics[trim=10 3 56 39, clip, width=\columnwidth]{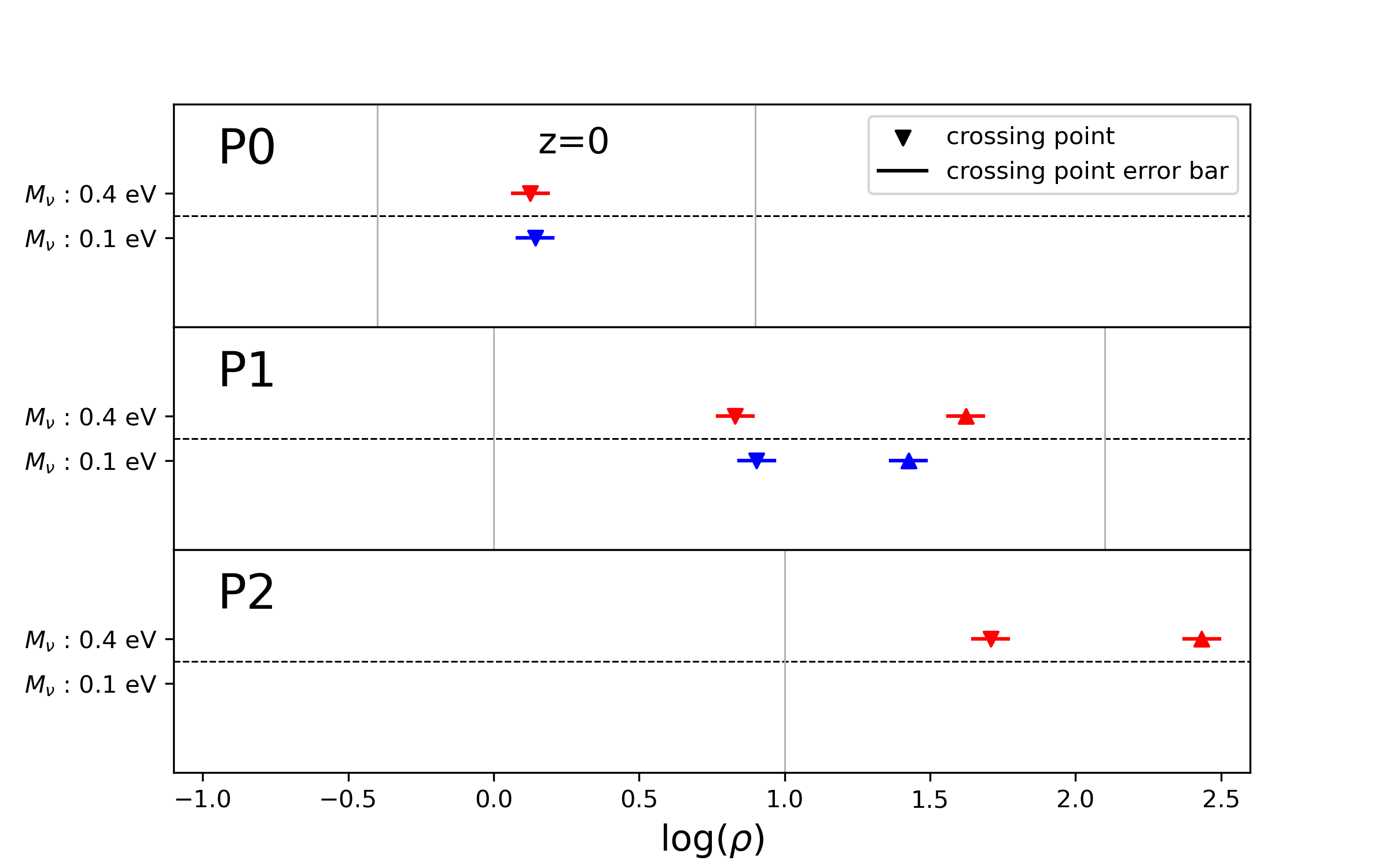}
\caption{Crossing points of the mean Betti curves for the \Quijote{} halo mass density fields. 
The figure layout mirrors that of \cref{fig:particles-betti-crossing}, with panels organized by redshift 
(\(z = \numlist{2;1;0}\)) and pair type (\(P_{0}\), \(P_{1}\), \(P_{2}\)).
Upward and downward (when present) crossings for the different massive neutrino models 
(\qty{0.1}{\electronvolt} in blue, \qty{0.4}{\electronvolt} in red) are indicated, with error bars
showing the standard deviation over \num{100} realizations per cosmology.}
\label{fig:halo-betti-crossing}
\end{figure}

Finally, we show the mean positions of the \Quijote{} halo Betti curve crossing points in \cref{fig:halo-betti-crossing}, 
the halo counterpart of \cref{fig:particles-betti-crossing}. The top, middle, and bottom panels correspond to 
\(z = \numlist{2;1;0}\), respectively. 
The horizontal axis indicates the density threshold, and each panel is subdivided into horizontal sub-panels 
displaying the crossing points for the different neutrino masses and pair types, as labeled.

As seen in \cref{fig:halos-betti}, at \(z=2\) 
each pair type exhibits a single upward crossing point, marked by an upward triangle. 
For this redshift, the crossing points of the \qty{0.4}{\electronvolt} massive neutrinos model occur at slightly higher densities 
than those of the \qty{0.1}{\electronvolt} model. However, when considering the error bars, the 
positions are statistically compatible. Across pair types \(P_{0}\) to \(P_{2}\),
the crossing points shift to increasingly higher densities, consistent with the behavior observed in the corresponding 
persistence diagrams and Betti curves.
At  \(z=1\), the upward crossing points are still present  for each pair type. In addition, a
new downward crossing appears for \(P_{1}\) 
on the lower-density side of the upward crossing, but only for the \qty{0.4}{\electronvolt} massive neutrino model.
At \(z=0\), the  \(P_{0}\) crossing that was upward at higher redshift becomes downward, with no additional crossings. 
For \(P_{1}\) pairs, both neutrino models exhibit a downward-upward pair of crossings, with the upward crossing occurring 
at a higher density threshold for the larger neutrino mass. 
For \(P_{2}\) pairs, a downward-upward crossing pair appears at 
\(z=0\)  only for the \qty{0.4}{\electronvolt} scenario, 
while no crossing is observed for the \qty{0.1}{\electronvolt} model.

Remarkably, as for the \gls{dm}  
field, our persistent homology analysis of halo density fields also reveals 
detectable neutrino mass-dependent signatures, 
particularly at high redshift, confirming that halos provide a 
complementary and robust probe of massive neutrinos in the cosmic web. 


\section{Conclusions and Outlook}      \label{sec:conclusions}

Traditional summary statistics such as the two-point correlation function and power spectrum compress the density field, 
encoding only pairwise correlations while discarding higher-order structural information. In contrast, 
persistence-based topological metrics 
retain the higher-order, \textit{multiscale} patterns of the cosmic web. 
Persistence diagrams, apexes of high-persistence features, 
and derived quantities such as Betti curves characterize 
the abundance and connectivity of topological features---independent islands, loops, and cavities---as a function of a filtration parameter.
Combined with classical statistics, these \textit{multiscale} topological measures provide a more 
complete description of the large-scale matter distribution, 
with the potential to break parameter degeneracies and tighten constraints 
on cosmological parameters, particularly the sum of neutrino masses, 
by capturing subtle cosmic web signatures that are otherwise inaccessible to conventional statistics.

In this work, building on our foundational studies of the impact of massive neutrinos on the 
\textit{multiscale} cosmic web via persistent homology \citep{Rossi:2022,Rossi:2026}, we quantified their imprint on persistence diagrams, 
Betti curves, and related topological statistics
using snapshots from the \Quijote{} $N$-body
simulations. We analyzed both the \gls{dm} density field on a regular grid and the corresponding 
halo catalogs, enabling a consistent comparison across tracers.
We considered three representative cosmologies sharing identical initial conditions and random seeds---ensuring that all 
morphological differences are driven by neutrino mass effects rather than sample variance or transients: a fiducial Planck-like 
massless neutrino model, and two massive neutrino scenarios with \({M_{\nu}=\qtylist{0.1;0.4}{\electronvolt}}\). 
We examined three redshift slices, \(z = \numlist{2;1;0}\), and in all cases results were averaged 
over \num{100} realizations per cosmology at each redshift.
For the \gls{dm} density field, we adopted a construction approach similar to \citet{Moon:2023}. 
For halo catalogs, the density field was built via Delaunay triangulation using \gls{dtfe}, suitable 
for the lower sampling density \citep{Rossi:2026}.

Our key findings are summarized as follows:
\begin{enumerate}
  \item {\bfseries Persistence diagrams:} For the \gls{dm} density field, characteristic triangular shapes are observed for all pair types
   (\(P_{0}\), \(P_{1}\), \(P_{2}\)) across \(z = \numlist{2;1;0}\), 
   with low-persistence bases and high-persistence apexes encoding cosmic-web transitions.
  Massive neutrinos induce systematic shifts in the diagrams: void-wall (\(P_{0}\)),  saddle-saddle (\(P_{1}\)),  
  and filament-peak (\(P_{2}\)) pairs show clear density-dependent differences, with \(P_{1}\)
  pairs particularly sensitive at high redshift. Depending on redshift and pair type, neutrino masses shift the diagrams toward higher or lower densities.
  These changes manifest in the corresponding Betti curves as higher values at low and high densities and lower values at intermediate densities.
  Halo persistence diagrams exhibit similar shapes and trends, though with less sharply defined edges and a disk-segment-like form for 
  \(P_{1}\)  pairs at \(z=0\). Despite the lower sampling, \(P_{1}\)  pairs at \(z=2\)  show the largest and most robust shifts, while 
  \(P_{0}\) and \(P_{2}\) pairs also display measurable shifts, confirming that topological features in halos are also sensitive probes of neutrino effects.
 
  \item {\bfseries Apexes:} High-persistence 
  apex positions in persistence diagrams are particularly sensitive to neutrino masses,
  especially for 
  \(P_{1}\) pairs at \(z=2\).
  In the \gls{dm} density field, apex positions are clearly separated between the 
  fiducial massless model and the heaviest neutrino case (\qty{0.4}{\electronvolt}) across all redshifts and pair types. 
  For \(P_{1}\) pairs, the apex shifts by more than \num{2}\(\sigma\) when moving from the fiducial model to the \qty{0.1}{\electronvolt}
  cosmology
  at  \(z = 1\) and \(z = 2\). 
  Halo apexes exhibit similar sensitivity than \gls{dm} density fields at \(z=2\), particularly for \(P_{1}\) and \(P_{2}\) pairs, 
  although the direction of the shift can differ from the \gls{dm} case.
  Differences decrease at lower redshifts, reflecting the evolving influence of neutrino free-streaming.
   
  \item {\bfseries Betti curves:} For the \gls{dm} density field, Betti curves broaden and flatten with increasing neutrino mass, 
  exhibiting two characteristic density thresholds where Betti numbers remain invariant.
  Massive neutrinos reduce the Betti numbers around the curve peaks at all redshifts, with the strongest effect for 
  \(P_{0}\) pairs (\(\approx\qty{2}{\percent}\) 
  for \qty{0.1}{\electronvolt},   \(\approx\qty{10}{\percent}\) for \qty{0.4}{\electronvolt}),
  intermediate for  \(P_{2}\) pairs (\(\approx\qty{1}{\percent}\)  and \(\approx\qty{5}{\percent}\)),
  and and weakest for  \(P_{1}\) pairs (<\(\qty{1}{\percent}\) and  \(\approx\qty{2}{\percent}\)), respectively.
  For halos, despite the sparser sampling, the Betti curves retain clear topological signatures 
  and enable robust cosmology comparisons.
  Differences are especially pronounced at \(z=2\), where Betti numbers at the peak 
  exceed the fiducial model by \qty{10}{\percent} for \qty{0.1}{\electronvolt}  
  and \qty{20}{\percent} for \qty{0.4}{\electronvolt}  neutrinos. 
  \(P_{2}\) pairs show the most distinct separation, followed by \(P_{1}\),
  while  \(P_{0}\) pairs exhibit larger uncertainties that reduce statistical significance.
  These trends mirror those observed in persistence diagrams, highlighting the robustness 
  of topological statistics even for sparse halo tracers.

  \item {\bfseries Betti crossings:} Crossing points of the Betti curves, defined as the densities where the ratio of 
  the massive neutrino Betti curve to the fiducial massless cosmology equals one, provide an interpretable, 
  complementary signature of neutrino effects, confirming trends 
  seen in persistence diagrams and apex shifts.
  In the \gls{dm} density field, crossings are largely insensitive to neutrino mass. For \(P_{0}\), \(P_{1}\), and \(P_{2}\),
 ``downward'' and ``upward'' crossings generally shift to higher densities, reflecting the ordering of the Betti-curve peaks:
  \(P_{0}\)  on the low-density side,  \(P_{2}\)  on the high-density side.
  Notably, the upward crossing of  \(P_{2}\) pairs moves to higher densities more rapidly with increasing neutrino mass, with the points clearly separated at 
  \(z=0\). For \(P_{0}\) pairs, an upward crossing is absent, occurring only near the Betti-curve edge where all curves converge to zero and uncertainties are large.
  In halo fields, the crossings capture the combined effect of neutrino suppression and non-linear halo evolution.
  At \(z=2\), all pair types show a single upward crossing, shifted to slightly higher densities for massive neutrinos. 
  As redshift decreases, crossings for \(P_{1}\) and  \(P_{2}\) evolve into downward-upward pairs, reflecting the growing influence of non-linear structure.
  Across all pair types, crossing points shift to increasingly higher densities, consistent with trends observed in the 
  corresponding persistence diagrams and Betti curves. 
  
\end{enumerate}

In particular, for the first time, we have demonstrated that high-persistence apex positions in persistence diagrams 
are sensitive to neutrino masses, especially for saddle statistics (\(P_{1}\) pairs)
at high redshift (\(z=2\)). 
Mass-dependent signatures from our persistent homology analysis, in both the \gls{dm} and halo fields, 
reveal detectable neutrino signatures at the few-percent level, even for $M_\nu \sim 0.1$ eV,
providing a robust, physically grounded probe of massive neutrinos in the cosmic web. 
These results thus establish a solid foundation for forward-modeling or emulator-based approaches (Michaux et al, in prep.) using persistent homology 
and environment-based metrics to constrain (or detect) neutrino mass along with additional cosmological parameters, 
with direct implications for ongoing and upcoming surveys (e.g., DESI, Euclid, Rubin-LSST). 

Building on our foundational studies \citep{Rossi:2022,Rossi:2026}, this work represents a further step toward quantifying the 
constraining power of topological statistics---both individually and in combination with classical measures---for robust inference of cosmological parameters, 
particularly the sum of neutrino masses. Beyond persistence diagrams and Betti curves, persistent homology and \gls{tda} enable the extraction of a broad 
set of complementary statistics, including filament-based measures. Characterizing the subtle neutrino signatures across these observables provides a 
crucial foundation for their combined and reliable use in parameter inference.

To this end, further work will be pursued to fully exploit the constraining power of persistence-based topological statistics. 
Our ongoing program, characterizing the impact of massive neutrinos on the \textit{multiscale} cosmic web via global topology and persistent homology, 
includes optimizing summary statistics and related normalization procedures, developing likelihood frameworks, 
and implementing machine-learning and  emulator strategies  (i.e., Michaux et al, in prep.)
to translate persistence-based observables 
into quantitative constraints on neutrino mass---including potential sensitivity to the mass hierarchy---and other cosmological parameters in extended cosmological scenarios. 
Parallel efforts address systematic observational effects (e.g., redshift-space distortions, survey incompleteness, and tracer bias) as well as the role of baryonic physics, 
with direct applications to current state-of-the-art survey data such as DESI. These efforts will provide a solid foundation for leveraging topological 
measures as robust cosmological probes in ongoing and future surveys.


\begin{acknowledgements}\label{sec:acknowledgements}

This work was supported by the National Research Foundation of Korea (NRF) grant funded by the Korea government (MSIT) (No. NRF-RS-2026-25472461). 
We also acknowledge the use of our computing resources at Sejong University (Xeon Silver 4114 and Xeon Gold 6126 node architectures).

\software{
  \DisPerSE{} \citep{DisPerSE:2011a,DisPerSE:2011b},
  Matplotlib \citep{Matplotlib},
  NumPy \citep{NumPy},
  \Pylians{} \citep{Pylians:2018},
  SciPy \citep{SciPy}.
}

\end{acknowledgements}


\bibliography{references}{}
\bibliographystyle{aasjournalv7}


\end{document}